\documentclass[11pt]{article}

\usepackage{amssymb,latexsym}
\usepackage{epsfig}
\usepackage{mathrsfs}
\usepackage{eufrak}
\newtheorem{theorem}{Theorem}
\newtheorem{lemma}{Lemma}

\newcommand{\be}{\begin{equation}}
\newcommand{\ee}{\end{equation}}
\newcommand{\pfbox}{\hfill\mbox{$\Box$}}
\newenvironment{pf}{\paragraph*{Proof{\rm.}}}{\pfbox\bigskip}

\begin{document}

\title{{\bf A New Family of Unitary Space-Time Codes with a Fast Parallel
Sphere Decoder Algorithm}
\thanks{The authors are with
Department of Electrical and Computer Engineering, Louisiana State
University, Baton Rouge, LA 70803; Email: \{chan, kemin,
aravena\}@ece.lsu.edu, Tel: (225)578-\{8961, 5533,5537\}, and Fax:
(225) 578-5200. }}

\author{Xinjia Chen, Kemin Zhou and Jorge Aravena}

\date{June 2007}

\maketitle

\begin{abstract}

In this paper we propose a new design criterion and a new class of
unitary signal constellations for differential space-time
modulation for multiple-antenna systems over Rayleigh flat-fading
channels with unknown fading coefficients. Extensive simulations
show that the new codes have significantly better performance than
existing codes. We have compared the performance of our codes with
differential detection schemes using orthogonal design, Cayley
differential codes, fixed-point-free group codes and product of
groups and for the same bit error rate, our codes allow smaller
signal to noise ratio by as much as 10 dB.

The design of the new codes is accomplished in a systematic way
through the optimization of a performance index that closely
describes the bit  error rate as a function of the signal to noise
ratio. The new performance index is computationally simple and we
have derived analytical expressions for its gradient with respect
to constellation parameters.

Decoding of the proposed constellations  is reduced to a set of
one-dimensional closest point problems that we solve using
parallel sphere decoder algorithms. This decoding strategy can
also improve efficiency of existing codes.

\end{abstract}

\section{Introduction}

Recently there have been extensive research interests in wireless
communication links with multiple transmitter antennas. For the
Rayleigh-fading channel models, information-theoretic analysis has
shown that the capacity of a communication link with multiple
transmitter antennas can substantially exceed that of a
single-antenna link \cite{Foschini1}, \cite{Foschini2},
\cite{Marzetta}, \cite{Teletar}, \cite{Zheng}. Several coding and
modulation schemes have also been proposed to exploit the potential
increase in the capacity through space diversity. For the coherent
multiple-antenna channel, several transmit diversity methods and
code construction have been presented in \cite{Alamouti},
\cite{Tarokh1}, \cite{Tarokh2} and the references therein (see,
e.g., \cite{Damen2},  \cite{Gamal}, \cite{Guey}, \cite{Hassibi2},
\cite{Narula}--\cite{Raleigh}, \cite{Seshadri}, \cite{Tirkkonen},
\cite{Winters}--\cite{Wittneben2}). In particular, Tarokh, Seshadri,
and Calderbank \cite{Tarokh1} proposed space-time codes which
combine signal processing at the receiver with coding techniques
appropriate to multiple transmitter antennas. Alamouti
\cite{Alamouti} discovered a remarkable transmitter diversity scheme
for two transmitter antennas, which was later generalized by Tarokh
{\it et al.} \cite{Tarokh2} as a framework for space-time block
codes. Motivated by the fact that, in many situations, channel state
information may not be available to the receiver, Hochwald and
Marzetta \cite{{Hochwald3}} proposed a general signaling scheme,
called unitary space-time modulation, and showed that this scheme
can achieve a high ratio of channel capacity in combination with
channel coding. The design of unitary space-time constellations was
investigated in \cite{Agrawal}, \cite{Hassibi3} and
\cite{Hochwald1}. More recently, differential modulation and code
construction methods for multiple transmit antennas have been
proposed by Hochwald {\it et al.} \cite{Hochwald2}, Hughes
\cite{Hughes1},  Tarokh {\it et al.} \cite{Jafarkani, Tarokh3} and
some other researchers (see, e.g., \cite{Borran}, \cite{Hassibi1},
\cite{Hughes2}, \cite{Hwang}, \cite{Liang}, \cite{Liu},
\cite{McCloud}, \cite{Tao, Tao2}, \cite{Warrier}, \cite{Xia}).

We investigate the encoding and decoding issues for the
differential unitary space-time modulation scheme independently
proposed by Hochwald and Sweldens in \cite{Hochwald2} and Hughes
in \cite{Hughes1}.  A number of unitary space-time codes have been
proposed aimed at achieving high performance, low encoding and
decoding complexity. Among these, we recall the orthogonal design
(see, \cite{Jafarkani, Tarokh3}), cyclic group codes
\cite{Hochwald2, Hughes1},  Caley differential (CD) codes
\cite{Hassibi1} and the {\it full-diversity} codes such as
fixed-point-free (FPF) unitary group codes $G_{m,r}$, non-group
codes $S_{m,s}$ and products of cyclic groups \cite{Shokrollahi}.
Orthogonal design has extremely low decoding complexity;
unfortunately, the performance degrades significantly when the
number of receiver antennas is more than one or the data rate is
high. Caley differential codes and the full-diversity codes
outperform orthogonal designs in many cases, while the decoding
complexity is much higher than that of orthogonal designs. The
main idea of decoding the full-diversity codes and Caley
differential codes is to formulate the decoding problem as a
closest point problem and then solve it by existing methods such
as ``LLL'' lattice algorithm and sphere decoder algorithm. The
decoding complexity depends critically on the dimension of the
underlying closest point problem.

In this paper we develop a new paradigm for the design of high
performance, low encoding and decoding complexity, unitary
space-time codes. Similar to the full-diversity codes $G_{m, r}$,
$S_{m,s}$ and products of cyclic groups, our proposed
constellations also use diagonal matrices as the kernel for fast
decoding purpose. However, in sharp contrast to those existing
codes which are parameterized by special integers, our
constellations are defined by real-valued parameters and are not
restricted to have {\it full diversity} or group structure.
Consequently, unitary space-time code with our proposed structure
exists for any combination of antennas and constellation size.

We define a code performance index that describes the bit error
rate as a function of the signal to noise ratio. The index is
simple to evaluate yet highly accurate in the normal signal to
noise ratio (SNR) region. As a result, it is possible to bring all
the power of non-linear programming into the code design. We have
developed a complete gradient descent algorithm to design
constellations that are optimal with respect to the bit error
rate.  It should be noted that the idea of code design by
gradient-based optimization for non-coherent MIMO channels was
proposed before in \cite{Agrawal} and \cite{Hassibi1}.  A
systematic design of unitary constellation based on random search
has been proposed in \cite{Hochwald1}.  Our approach differs from
the previous works in the design criterion, the structure of
signal constellations, and the decoding method. We attempt to
apply gradient descent techniques to directly minimize the bit
error rate over signal constellations which allow for efficient
decoding algorithms.

Exploiting the special structure of our proposed constellations, the decoding problem is reduced to
one-dimensional closest point problems which can be efficiently solved in parallel. Based on that strategy, we
have developed parallel sphere decoder algorithms which can also be applied to improve decoding efficiency of
existing codes.

Based on the new structure and using the optimal design
techniques, we have obtained constellations which significantly
outperform existing ones. For example, with spectral efficiency
$R=6$ bits per channel use, we have found a constellation which
improves upon orthogonal design by about $10$ dB at block error
rate $6 \times 10^{-2}$ when using two transmitter and receiver
antennas.  With the same configuration, the corresponding
improvement upon Caley differential code is about $9$ dB.

In the rest of this section we establish the notation and describe
the channel model. Section \ref{model} introduces the  structure
of the new constellations and develops the optimization procedure
for their design. Specifically, we introduce the performance index
that converts constellation design into a minimization problem
amenable to steepest descent techniques and derive simple
expressions for the computation of its gradient. Section
\ref{decoding} develops a parallel sphere decoder algorithm that
can also be applied to improve existing codes. Section
\ref{examples} presents results of the performed simulations.
Section \ref{conclusion} summarizes our findings. Proofs and
constellation data are provided in the Appendices.

\subsection{Notation}

Throughout this paper, we use the following notations.

$\mathbb{R}$ --- real number field;

$\mathbb{C}$ --- complex number field;

$\mathbb{Z}$ --- integer set;

$\lfloor . \rfloor$ --- floor function;

$\lceil . \rceil$ --- ceiling function;

$\lfloor x \rceil$ --- the integer closest to $x$;

${\rm mod}^*(x)$ --- symmetric modulus operation such that ${\rm
mod}^*(x)$ has range $[ - \frac{x}{2}, \frac{x}{2})$;

$\arg(.)$ --- phase angle operator taking values in $[ -\pi , \;
\pi)$;

$\det(.)$ --- determinant function;

${\rm tr}(.)$ --- trace function;

${\rm diag}([x_1, \cdots,x_n])$ --- diagonal matrix with $x_p$ at
the $p$-th row and the $p$-th column;

$||X||$ --- Euclidean norm of vector $X$;

$||X||_\mathrm{F}$ --- Frobenius norm of matrix $X$;

$[X]_{pq}$ --- entry of $X$ at the $p$-th row and $q$-th column;

$\Re(X)$ --- real part of $X$;

$\Im(X)$ --- imaginary part of $X$;

$X^\intercal$ --- transpose of $X$;

$X^\dag$ --- conjugate transpose of $X$;

${\rm abs}(X)$ --- the matrix obtained by replacing each entry of
$X$ with its modulus;

$\mathcal{ CN}(0,1)$ --- complex random variable with zero mean and variance one.

$\nabla g({\bf x})$ --- gradient of function $g({\bf x})$.

\subsection{Channel Model}

Consider a communication link with $M$ transmitter and $N$
receiver antennas operating in a Rayleigh flat-fading channel,
which can be described by the following channel model
\cite{Hochwald2}
\[
X_{\tau} = \sqrt{\rho} S_{\tau} H_{\tau} + W_{\tau}
\]
where $\tau$ is the index of time frame, $H_{\tau} \in \mathbb{C}^{M \times N}$ is the channel matrix with
$\mathcal{ CN}(0,1)$ entries and is unknown to the receiver and the transmitter, $S_{\tau} \in \mathbb{C}^{M
\times M}$ is the transmitted signal, $X_{\tau} \in \mathbb{C}^{M \times N}$ is the received signal, $W_{\tau}
\in \mathbb{C}^{M \times N}$ is Gaussian noise with $\mathcal{ CN}(0,1)$ entries, and $\rho$ is the expected SNR
at each receiver antenna. It should be noted that the channel matrix $H_\tau$ has been normalized so that the
SNR is not dependent on the number of transmitter antennas.  It is assumed that the channel matrix is
approximately constant within two consecutive time frames, i.e., $H_\tau \approx H_{\tau -1}$. However, for the
$\tau$-th and the $\iota$-th time frames that are not consecutive, $H_\tau$ and $H_{\iota}$ are mutually
independent and thus their realizations can be significantly different.  The transmitted signals are determined
by the following fundamental differential transmitter equations \cite{Hochwald2}
\[
S_0 = I_{M \times M}, \;\;\;\; S_{\tau} = V_{\tau} S_{\tau-1}, \;
\tau = 1, 2, \cdots
\]
where $V_{\tau} \in \mathbb{C}^{M \times M}$ is a unitary matrix picked from signal constellation $\mathcal{
V}$. It is shown in \cite{Hochwald2, Hochwald3, Hughes1} that the maximum-likelihood (ML) detection is to
minimize
\[
||X_{\tau} - V_\ell X_{\tau-1}||_{\mathrm{F}}^2
\]
among all possible $V_\ell \in \mathcal{ V}$.   The Chernoff bound of pair-wise probability of mistaking
$V_{\ell}$ for $V_{\ell^{'}}$ or vice versa is given by \cite{{Hochwald3}} \be P (V_\ell, \; V_{\ell^{'}}) =
\frac{1}{2} \prod_{m=1}^M \left[ 1 + \frac{\rho^2 \sigma_m^2 }{4( 1 + 2 \rho) } \right]^{-N} \label{chernoff}
\ee where $\sigma_m$ is the $m$-th singular value of $V_\ell - V_{\ell^{'}}$.

\section{A New Constellation Design Approach} \label{model}

The new code design paradigm that we propose uses diagonal
matrices, as in \cite{Shokrollahi}, to simplify the decoding
process.  Our approach is similar to \cite{Agrawal} and
\cite{Hassibi1} in the spirit of relaxing the code structures from
strict structures such as orthogonal or diagonal structure,
parameterizing the codes,  and employing the powerful
gradient-based optimization to find the best codes.  A significant
new feature is the ability to formulate the design as a non-linear
programming problem that directly minimizes the bit error rate. In
the following, we begin our presentation by introducing new codes
that are functions of real-valued variables and do not require
full diversity. Then we introduce the cost function and derive
expressions for its gradient that are used in a steepest descent
design algorithm.

\subsection{Constellation Structure}
In this section, we introduce a new class of unitary space-time
codes which can be efficiently encoded and decoded. Similar to the
full-diversity codes such as FPF code $G_{m, r}$, non-group code
$S_{m,s}$ and products of cyclic groups \cite{Shokrollahi}, our
proposed constellation also involves diagonal matrices for fast
decoding purpose. However, in sharp contrast to those
full-diversity codes which are parameterized by particular
integers, our proposed constellations are determined by continuous
parameters and are not restricted to have full diversity or group
structure. Consequently, unitary space-time code with our proposed
structure exists for any combination of antennas and constellation
size.

 Let $b \geq 0$ be an integer and let $L$ be a power of $2$.
 We construct a constellation $\mathcal{ V}$ with $\mathscr{L} = 2^b L$ signal matrices as
follows.

For $q = 0, 1, \; \cdots, \;2^b - 1$, define
\[
\Lambda_q \stackrel{\mathrm{def}}{=} {\rm diag} \left( \left[\exp \left( \frac{i 2 \pi \lambda_{q,1}}{L}
\right), \; \cdots, \; \exp \left( \frac{i 2 \pi \lambda_{q,M}}{L} \right ) \right] \right),
\]
where $\lambda_{q,1} = 1$ and $\lambda_{q,m} \in [0, \; L), \;\; m
= 2, \cdots, M$ are {\it real-valued} parameters. Let $A_0 = B_0 =
I$ and $A_q, \; B_q, \;\; q = 1, \; \cdots, \;2^b - 1$ be unitary
matrices. Then the constellation is given by
\[
\mathcal{ V} = \{A_q \Lambda_q^\ell B_q \;\; | \;\; \ell =0, \; 1, \; \cdots, \; L-1; \;\; q = 0, 1, \; \cdots,
\; 2^b - 1  \}.
\]

We note that the constellation design problem is to find
$\lambda_{q,m}$ and $A_q, \;\;B_q$ so that the bit error rate  is
minimized. We shall show that this problem can be solved
efficiently.

For the purpose of comparing our constellations with existing
ones, we note that the spectral efficiency of our proposed
constellation is
\[
R = \frac{ \log_2 (\mathscr{L}) }{M} = \frac{ b + \log_2 (L) }
{M}.
\]

It should be noted that, for the special case $b = 0$, the signal
constellation reduces to
\[
\{ \Lambda^\ell \; | \; \ell = 0, 1, \cdots, L-1\}
\]
where
\[
\Lambda = {\rm diag} \left( \left[\exp \left( \frac{i 2 \pi
\lambda_{1}}{L} \right), \; \cdots, \; \exp \left( \frac{i 2 \pi
\lambda_{M}}{L} \right ) \right] \right)
\]
with $\lambda_{1} = 1$ and {\it continuous} parameters
$\lambda_{m} \in [0, \; L), \;\; m = 2, \cdots, M$. We refer to
such constellation as a {\it continuous diagonal code}. Obviously,
it is a generalization of cyclic group code.

In general, with fixed constellation size $\mathscr{L}$, the
performance may be significantly improved by increasing the number
of blocks (i.e., $\frac{\mathscr{L}} {L}$). Interestingly, we
shall show that the decoding complexity increases slightly with
respect to the number of blocks.  This property can be attributed
to our
 parallel sphere decoder algorithms, discussed in Section \ref{decoding}.

\subsection{Design Performance Index}

Efficient constellation design is a challenging task due to the large number of parameters.  In addition to the
structure of the constellations, the design criterion is also critical for the achievable bit error rate
performance.  One of the widely used criterion is to use the diversity product as the performance measure of a
constellation.  The design objective is to maximize the diversity product over a class of constellations that
have full diversity (see, e.g., \cite{Hochwald2} \cite{Liang}, \cite{Shokrollahi} and the references therein).
The drawbacks of the conventional design criterion are the following:  First, the diversity product is
essentially a worst-case measure. In many situations, the overall performance of a constellation is not governed
by the behavior of extreme signal matrices.  As can be seen from our experimental results in Section
\ref{examples}, it is not uncommon to have constellations with {\it zero diversity product} significantly
outperforming constellations with the largest diversity product previously known.   Second, the measure
diversity product is derived by an asymptotic argument. The idea is that, as the SNR tends to infinity, the
Chernoff bound of the pair-wise error probability is dominated by the determinant of the difference of the pair
of unitary matrices. Such asymptotic argument is not flawless.  It is not clear how large the value of SNR can
be approximated as infinity so that no significant inaccuracy will be introduced in the evaluation of the block
(or bit) error rate.

In light of the limitations of the worst-case and asymptotic design criterion, we have established a new design
criterion which incorporates the bits assignment in the optimization of constellations.   Instead of using a
worst-case criterion such as diversity product \cite{Hochwald2}, we introduce a performance index that measures
directly the bit error rate as a function of the signal to noise ratio. The index is analytically tractable and
possesses simple analytical expressions for its gradient. Motivated by the fact that, for large constellation
size, the bit error rate may not be well governed by the block error rate, we shall also incorporate the bit
assignments in the process of constellation optimization. In particular we propose the cost function
\[
J=\int_{\rho_1}^{\rho_2} \log_{10} P_\mathrm{bit} (\rho) \; d
\log_{10}(\rho)
\]
where $P_\mathrm{bit} (\rho)$ is the union bound of bit error
probability and $[\rho_1, \rho_2]$ is the interval of SNR of
practical interests. We shall show that this cost function can be
well approximated by a very simple analytical expression.

\begin{figure}
\centering
\includegraphics[width=3.3in]{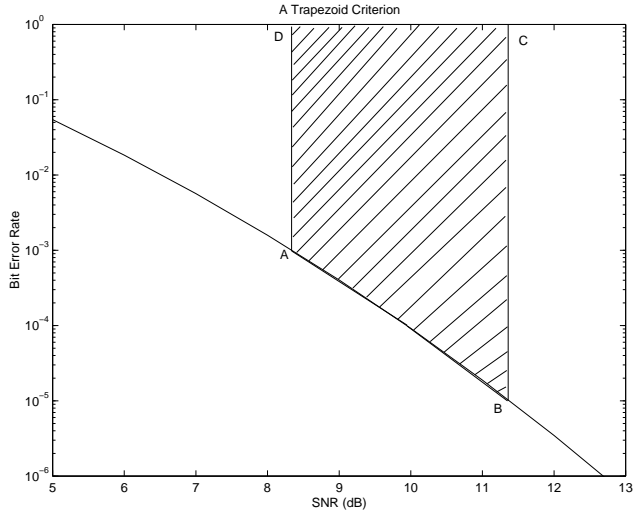}
\caption{The area of trapezoid ABCD, or equivalently $- \zeta(\mathcal{ V})$, reflects the quality of
constellation $\mathcal{ V}$.} \label{fig00}
\end{figure}

From numerous simulation results published in the literature, we
notice that, on a log scale, the bit error rate is an almost
linear function of the SNR.  Such phenomenon can be illustrated by
making use of the Chernoff bound (\ref{chernoff}).  For large SNR,
the Chernoff bound $P(V_\ell,V_{\ell^{'}})$ of pair-wise error
probability can be approximated by
\begin{eqnarray*}
P(V_\ell,V_{\ell^{'}}) & \approx & \frac{1}{2} \rho^{-MN} \left(
\prod_{m=1}^M
\frac{\sigma_m^2 }{8} \right)^{-N}\\
& = & \frac{1}{2} \rho^{-MN} \left( \frac{ \det(V_\ell -
V_{\ell^{'}}) }{8} \right)^{-N}.
\end{eqnarray*}
Since such approximation is tight for most combinations $(\ell,
\ell^\prime)$ and $V_\ell$ is assumed to be equally likely for all
$\ell$, the union bound of the bit error rate is well approximated
by
\[
P_\mathrm{bit} (\rho) \approx \rho^{-MN} \; \frac{ \sum_{\ell \neq
\ell^\prime} \;d^\mathrm{H}(\ell, \ell^\prime) \left( \frac{
\det(V_\ell - V_{\ell^{'}}) }{8} \right)^{-N}} {2 \mathscr{L}
\log_2 \mathscr{L}}
\]
where $d^\mathrm{H}(\ell, \ell^\prime)$ denotes the Hamming
distance of bits assigned to $V_\ell$ and $V_{\ell^\prime}$.
Applying logarithm operation gives
\begin{eqnarray*}
&   & \log_{10} P_\mathrm{bit} (\rho)\\
 & \approx & - \frac{MN}{10}  \; 10 \log_{10} (\rho)\\
&   &   + \log_{10} \left( \frac{ \sum_{\ell \neq \ell^\prime} \; d^\mathrm{H}(\ell, \ell^\prime) \; \left(
\frac{ \det( V_\ell - V_{\ell^\prime} ) }{8} \right)^{-N} } {2 \mathscr{L} \log_2 \mathscr{L} } \right).
\end{eqnarray*}
Figure \ref{fig00} displays the actual cost function and the
proposed approximation. For completeness we mention that the block
error rate admits a similar approximation.

Due to the excellent linearity of the performance curve in the
logarithm scale, the cost function can be well approximated by
\begin{eqnarray*}
\zeta(\mathcal{ V}) & \stackrel{\mathrm{def}}{=} &  [ \log_{10} P_\mathrm{bit} (\rho_{2}) + \log_{10}
P_\mathrm{bit}
(\rho_{1}) ]\\
&   & \times \; [ \log_{10}(\rho_{2}) - \log_{10}(\rho_1) ].
\end{eqnarray*}
We propose to design constellations that minimize the index $\zeta(\mathcal{ V})$.

In practice, we can choose $\rho_1$ and $\rho_2$ based on the
performance of the best cyclic group codes previously known. More
specifically, $\rho_1$ and $\rho_2$ can be selected so that two
typical levels of bit error rate are respectively guaranteed. For
example, we can find $\rho_1$ and $\rho_2$ such that
\[
\log_{10} P_\mathrm{bit} (\rho_{1}) = 10^{-3}, \;\;\;\log_{10}
P_\mathrm{bit} (\rho_{2}) = 10^{-5}
\]
by a bisection method for an existing cyclic group code. When $\rho_1$ and $\rho_2$ have been found, the
criterion measure $\zeta(\mathcal{ V})$ is SNR independent. Most importantly,  the gradient of $\zeta(\mathcal{
V})$ with respect to the constellation parameters can be computed efficiently and thus allows for a gradient
descent method for constellation design. The optimization technique is described in the next section.

\subsection{Constellation Optimization}

In this section, we perform a global optimization to find unitary constellations of good performance. Our
strategy is to first choose the bit assignment and then search the code parameters to minimize the bit error
rate.  The advantage of this strategy is that the objective function $\zeta(\mathcal{ V})$ is a differentiable
function and is amenable for gradient-based optimization.  On the other hand, if we first search the good code
matrices and then try to find the best bit assignment, we need to solve a combinatorial optimization problem. In
general, such combinatorial problem is not tractable for gradient-based optimization techniques because the
objective function is not continuous. The only method for solving such combinatorial problem is the exhaustive
random search. Unfortunately, for large constellations, the searching can be extremely inefficient.

\subsubsection{Parameterization of Unitary Matrix}
In order to develop a gradient-based method for the minimization of the performance measure $\zeta(\mathcal{
V})$, the first step is to choose a suitable parameterization for unitary matrices.  The application of unitary
matrices parameterization \cite{Murnaghan} in signal constellation design has been pioneered by \cite{Agrawal}.
We adopt such idea of using parameterized unitary code matrices.  In general, a $M \times M$ unitary matrix $U$
can be determined by a set of $M^2$ parameters $\Theta$ defined as follows:
\begin{eqnarray*}
&   & \phi_{pq} \in \left[ -\frac{\pi}{2}, \frac{\pi}{2} \right],  \qquad 1 \leq p < q
  \leq M-1;\\
&   & \phi_{pM} \in [-\pi, \pi), \qquad 1 \leq p \leq M-1;\\
&   & \nu_{pq} \in \left[ -\frac{\pi}{2}, \frac{\pi}{2} \right], \qquad 1 \leq p < q  \leq M;\\
&   & \theta_k \in \left[ -\frac{\pi}{2}, \frac{\pi}{2} \right], \qquad k = 1, \cdots, M-1;\\
&   & \theta_M \in [-\pi, \pi).
\end{eqnarray*}

More specifically, let $U^{p,q}(\phi_{pq}, \nu_{pq})$ denote a $(M - p + 1)$-dimensional unitary matrix such
that
\begin{eqnarray*}
&   & [U^{p,q}]_{jk}\\
& = & \left\{\begin{array}{ll}
1, &  {\rm if}\; j = k \; {\rm and} \;  j \notin \{1,  q-p+1\}\\
\cos( \phi_{pq}  ), &  {\rm if}\; j = k \; {\rm and} \;  j \in \{1,
q-p+1\}\\
- \sin( \phi_{pq}  ) e^{-i \nu_{pq}}, &  {\rm if}\; j = 1 \; {\rm
and} \;  k = q-p+1\\
\sin( \phi_{pq}  ) e^{i \nu_{pq}}, &  {\rm if}\; k = 1 \; {\rm
and} \;  j = q-p+1\\
0, &  {\rm otherwise}
\end{array} \right.
\end{eqnarray*}
and let
\[
\digamma^r = U^{r, r+ 1} \; U^{r,r+2} \; \cdots \; U^{r,M},
\]
then, any unitary matrix $U(\Theta)$ can be represented as
\[
U = \mathscr{U}^{M-1} \; \digamma^{1}
\]
where
\[
\mathscr{U}^{1} = \left[\begin{array}{ll}
\exp(i \theta_{M-1}) & 0\\
0 & \exp(i \theta_{M}) \end{array}\right]
\]
and
\[
\mathscr{U}^{k + 1}  =  \left[\begin{array}{ll}
\exp(i \theta_{M-k-1}) & 0\\
0 & \mathscr{U}^{k} \; \digamma^{M-k}
\end{array}\right]
\]
for $k = 1, \cdots, M - 2$.

\subsubsection{Gradient Method}
Here we develop explicit expressions for the gradient of the performance measure $\zeta(\mathcal{ V})$. For the
computation of $\zeta(\mathcal{ V})$ we need to evaluate the union bound of the bit error rate, which depends on
the bit assignment.  With regard to the bit assignment, our intuition is that, if we first search the good code
matrices and then try to find the best bit pattern -- code matrix assignment, we need to cope with a
combinatorial optimization problem. Such combinatorial problem is generally not tractable for gradient-based
optimization techniques because the objective function is not differentiable. The available method for solving
such combinatorial problem will be random search. Unfortunately, for large constellations, the searching can be
extremely difficult. In our design, we shall first fix the bit assignment and then search the code parameters to
minimize the bit error rate.  In this way, the objective function is a differentiable function and is amenable
for gradient-based optimization techniques.

For simplicity, we use the binary-to-decimal conversion mapping scheme.  In such a scheme, a block of $b +
\log_2(L)$ bits is mapped into a signal matrix $A_q \Lambda_q^{\ell} B_q$ such that the first $b$ bits are the
binary representation of the block index $q$ and the remaining bits are the binary representation of the
diagonal index $\ell$. Let $d^\mathrm{H}(p, q, \ell, \ell^{'})$ denote the Hamming distance between the bits
respectively assigned to signal matrices $A_p \Lambda_p^{\ell} B_p$ and $A_q \Lambda_q^{\ell^{'}} B_q$. The
union bound of the bit error probability is then given by \be P_\mathrm{bit}  =   \frac{2}{2^b \; L \; [b +
\log_2(L)]}  \left ( \widehat{\mathcal{P}} + \widetilde{\mathcal{P}} \right ) \label{eqbit} \ee where
\[
\widehat{\mathcal{P}} = \sum_{p =0}^{2^b - 1} \; \sum_{\ell =0}^{L - 2} \; \sum_{\ell^{'} = \ell+1}^{L - 1}
d^\mathrm{H}(p, p, \ell, \ell^{'}) \; P( \Lambda_p^{\ell}, \; \Lambda_p^{\ell^{'}  } )
\]
and
\begin{eqnarray*}
&   & \widetilde{\mathcal{P}}\\
& = & \sum_{p =0}^{2^b - 2} \sum_{q = p+1}^{2^b - 1} \sum_{ \ell = 0}^{  L - 1  } \sum_{ \ell^{'} = 0}^{ L - 1 }
d^\mathrm{H}(p, q, \ell, \ell^{'}) P( A_p \Lambda_p^{\ell} B_p, A_q \Lambda_q^{\ell^{'}} B_q)
\end{eqnarray*}
It can be seen that, using (\ref{eqbit}) to compute $P_\mathrm{bit}$, the number of pair-wise error
probabilities to be evaluated is
\[
2^{b-1} L (L-1) + 2^{b-1} (2^b-1) L^2.
\]
The problem can still be solved using steepest descent method for
small to moderate $L$. However, the computational complexity may
be high for large $L$. For proof of concept, we focus here on the
special case of $\Lambda_q = \Lambda, \;\; A_q = I$ for $q = 0,
\cdots, 2^b -1$, where, exploiting the special structure of the
constellation, the number of pair-wise error probabilities to be
computed can be substantially reduced to
\[
(L-1) + 2^{b-1} (2^b -1) (2L-1).
\]
For this case, we have

\bigskip

\begin{theorem} \label{th_diag}
Let $d^\mathrm{H} (p,q)$ denote the Hamming distance between the
binary representation of integers $p$ and $q$.  Define
\[
w(k) = \sum_{\ell = 0}^{L-k-1} d^\mathrm{H}(\ell +k, \; \ell),
\;\; \;\; k = 0, 1, \cdots, L-1.
\]
Then
\[
P_\mathrm{bit} = \frac{2}{L [b + \log_2(L)]} \; \left[ \mathcal{P}^\prime + \sum_{k=1}^{L-1} w(k) \; P(I,
\Lambda^k) \right]
\]
where
\[
\mathcal{P}^\prime = \frac{1}{2^b} \sum_{p =0}^{2^b - 2} \sum_{q = p+1}^{2^b - 1} \sum_{ k = - L + 1  }^{ L - 1
}  [ w(|k|) + d^\mathrm{H} (p,q) ] P(B_p,\Lambda^k B_q).
\]
\end{theorem}

\bigskip

See Appendix \ref{app1} for a proof.  It should be noted that, to reduce computation, $w(k)$ can be pre-computed
and saved as a lookup table.

In order to use gradient descent method to minimize $\zeta(\mathcal{ V})$, we need to find the fastest descent
direction at every step of searching. Following the procedure in \cite{Agrawal}, we update $B_p$ as $B_p
U(\Theta)$.  In the sequel, we shall show that the computation of the gradient of performance measure
$\zeta(\mathcal{ V})$ reduces to the computation of: (i) the partial derivatives of functions of the form
$P(U(\Theta), \Phi)$ with respect to $\Theta$ at $\Theta = 0$ (i.e., all elements of $\Theta$ are zero); (ii)
the partial derivatives of functions of the form $P(\Lambda^{\ell}, \Phi)$ with respect to $\Lambda = {\rm diag}
( [e^ {2 \pi i \lambda_1 \slash L}, \cdots, e^ {2 \pi i \lambda_M \slash L}]  )$ at $\Lambda = I$ (i.e.,
$\lambda_m = 0, \; m = 1, \cdots, M$).

We have derived surprisingly simple formulas for computing
pair-wise error probabilities  and the related partial
derivatives.

\bigskip

\begin{theorem} \label{them2}
Let $U(\Theta)$ be unitary matrix parameterized by $\Theta$. Let $\Phi$ be a unitary matrix.  Let
\[
\alpha = \frac{ 4 ( 1 + 2 \rho)  } {  \rho^2 }, \quad \mathcal{Q} =  [ (\alpha + 2) I - \Phi - \Phi^\dag ]^{-1}
\Phi \] and
\[
\Lambda = {\rm diag} ( [e^ {2 \pi i \lambda_1 \slash L}, \cdots, e^ {2 \pi i \lambda_M \slash L}]  ). \]
 Then
\begin{eqnarray}
&   & P (I,  \Phi)  = \frac{ \alpha^{MN}} {2 \left( \det[ (\alpha + 2) I - \Phi - \Phi^\dag  ]  \right)^N },
\label{union_0}\\
&   & {\left. \frac{ \partial P( U, \Phi)  } { \partial \phi_{pq}  } \right|}_{\Theta = 0} = 2 N P( I,  \Phi) \;
\Re ([\mathcal{Q}]_{q p} - [\mathcal{Q}]_{p q}), \qquad \label{grad_0}\\
&   & {\left. \frac{ \partial P( U, \Phi) } { \partial \theta_{k}  } \right|}_{\Theta = 0} =  2 N P( I,  \Phi)
\; \Im ( [\mathcal{Q}]_{k k} ), \label{grad_1}\\
&   & {\left. \frac{ \partial P( U, \Phi)  } { \partial \nu_{pq}  } \right|}_{\Theta = 0} = 0, \label{grad_2}\\
&   & {\left. \frac{ \partial P(\Lambda^{\ell}, \Phi)  } { \partial \lambda_{m} } \right|}_{\Lambda = I} =
\frac{4 \pi  N \ell}{L} P(I, \Phi) \; \Im ( [\mathcal{Q}]_{mm} ). \label{grad_3}
\end{eqnarray}
\end{theorem}

\bigskip

See Appendix \ref{app2} for a proof. At the first glance, it is not clear how Theorem \ref{them2} can be applied
to the optimization of code matrices. From the expression of our performance metric $\zeta(\mathcal{V})$, it can
be seen that it suffices to compute the gradients of $P_\mathrm{bit} (\rho_{1})$ and $P_\mathrm{bit} (\rho_{2})$
with respect to code parameters. From (\ref{eqbit}), we can see that, since the bits assignment is fixed, it
suffices to compute the gradient of $P ( \Lambda_p^{\ell}, \; \Lambda_p^{\ell^{'}  } )$ and $P ( A_p
\Lambda_p^{\ell} B_p, \; A_q \Lambda_q^{\ell^{'}} B_q )$ with respect to code parameters for all combinations of
$p$ and $q$. Note that the first quantity can be viewed as a special case of the second. Hence, we focus on the
second quantity $P ( A_p \Lambda_p^{\ell} B_p, \; A_q \Lambda_q^{\ell^{'}} B_q  )$. We first consider how to
update the matrix $B_p$. In order to apply the gradient descent method to minimize the performance metric, we
need the partial derivatives of the function $P ( A_p \Lambda_p^{\ell} B_p, \; A_q \Lambda_q^{\ell^{'}} B_q )$
with respect to the parameters of $B_p$.  It can be seen from the complexity of the function $P(., .)$ and the
parameterization of $B_p$ that the direct computation of the partial derivatives can be extremely difficult.
Observing that, at every step, $B_p$ is to be updated as $\widehat{B}_p$ which is also a unitary matrix. Hence,
there must be an unitary matrix $U(\Theta)$ such that $\widehat{B}_p = B_p U(\Theta)$.  This means that we can
update the unitary matrices in a multiplicative way. As mentioned earlier, this method of updating unitary
matrices was proposed in \cite{Agrawal}.  In the same sprit with that of the conventional steepest-descent
minimization, to make $P ( A_p \Lambda_p^{\ell} B_p, \; A_q \Lambda_q^{\ell^{'}} B_q )$ descent in a fastest way
as $B_p$ is varying to a new matrix, we can choose $U(\Theta)$ based on the partial derivatives of $P ( A_p
\Lambda_p^{\ell} B_p U(\Theta), \; A_q \Lambda_q^{\ell^{'}} B_q )$ with respect to $\Theta$ at $\Theta = 0$. The
computation of the derivatives can be accomplished by applying Theorem \ref{them2} and the following fact:

{\it $P(., .)$ is invariant under unitary transforms.  That is,
for any unitary matrices $X$ and $Y$,  $P(U_L X U_R, \; Y) = P (X,
\;U_L^\dag Y U_R^\dag)$ for any unitary matrices $U_L$ and $U_R$.}

To prove this fact, we can use equation (\ref{identity}), which is shown in Appendix \ref{app2}.  By
(\ref{identity}),
\begin{eqnarray*} &   &  P (U_L X U_R, Y)\\
 & = & \frac{\alpha^{MN}} {2 \left( \det[ \alpha I + (U_L X
U_R - Y) (U_L X U_R - Y)^{\dag}  ] \right)^N }.
\end{eqnarray*}
Observing that
\begin{eqnarray*}
&   & (U_L X U_R - Y) (U_L X U_R - Y)^{\dag}\\
& = & U_L(X - U_L^\dag Y U_R^\dag)
 U_R U_R^\dag (X - U_L^\dag Y U_R^\dag)^{\dag} U_L^\dag\\
  & = & U_L(X - U_L^\dag Y U_R^\dag) (X - U_L^\dag Y U_R^\dag)^{\dag} U_L^\dag, \end{eqnarray*}
we have
\begin{eqnarray*}
&   & \alpha I + (U_L X U_R - Y) (U_L X U_R - Y)^{\dag}\\
& = & U_L \left [ \alpha I + (X - U_L^\dag Y U_R^\dag) (X - U_L^\dag Y U_R^\dag)^{\dag} \right ] U_L^\dag.
\end{eqnarray*}
Hence
\begin{eqnarray*}
&   & \det \left[ \alpha I + (U_L X U_R - Y) (U_L X U_R - Y)^{\dag} \right ]\\
& = & \det( U_L U_L^\dag) \det \left [ \alpha I + (X - U_L^\dag Y U_R^\dag) (X - U_L^\dag Y U_R^\dag)^{\dag}
\right ]\\
& = & \det \left [ \alpha I + (X - U_L^\dag Y U_R^\dag) (X - U_L^\dag Y U_R^\dag)^{\dag} \right ]
\end{eqnarray*}
and
\begin{eqnarray*} &   &  P (U_L X U_R, Y)\\
& = & \frac{\alpha^{MN}} {2 \left \{ \det \left [ \alpha I + (X - U_L^\dag Y U_R^\dag)
(X - U_L^\dag Y U_R^\dag)^{\dag}  \right ] \right \}^N }\\
& = & P(X, \; U_L^\dag Y U_R^\dag).
\end{eqnarray*}
This proves the invariant property.   An immediate result from such property is \[ P \left ( A_p
\Lambda_p^{\ell} B_p, \; A_q \Lambda_q^{\ell^{'}} B_q \right ) = P \left ( I, \; ( A_p \Lambda_p^{\ell}
B_p)^\dag A_q \Lambda_q^{\ell^{'}} B_q \right ),
\]
which implies that we can let $\Phi = ( A_p \Lambda_p^{\ell}
B_p)^\dag A_q \Lambda_q^{\ell^{'}} B_q$ and apply (\ref{union_0})
to compute $P( A_p \Lambda_p^{\ell} B_p, \; A_q
\Lambda_q^{\ell^{'}} B_q )$.

Making use of such property, we have
\begin{eqnarray*}
&   & P \left ( A_p \Lambda_p^{\ell} B_p U(\Theta), \; A_q \Lambda_q^{\ell^{'}} B_q \right )\\
& = & P \left ( U(\Theta), \; ( A_p \Lambda_p^{\ell} B_p)^\dag A_q \Lambda_q^{\ell^{'}} B_q \right ).
\end{eqnarray*} If we identify $( A_p \Lambda_p^{\ell} B_p)^\dag A_q \Lambda_q^{\ell^{'}} B_q$ as $\Phi$, we have
\[
P \left ( A_p \Lambda_p^{\ell} B_p U(\Theta), \; A_q \Lambda_q^{\ell^{'}} B_q \right ) = P (U(\Theta), \Phi).
\]
Hence, the partial derivatives of $P( A_p \Lambda_p^{\ell} B_p
U(\Theta), \; A_q \Lambda_q^{\ell^{'}} B_q )$ can be computed by
applying Theorem \ref{them2}.

Similarly, we can update $A_p$ as $A_p U(\Theta)$ and compute the
partial derivatives of
\begin{eqnarray*}
&   & P \left ( A_p U(\Theta) \Lambda_p^{\ell} B_p, \; A_q \Lambda_q^{\ell^{'}} B_q \right )\\
& = & P \left ( U(\Theta), \; A_p^\dag A_q \Lambda_q^{\ell^{'}} B_q (\Lambda_p^{\ell} B_p)^\dag \right )
\end{eqnarray*} with respect to $\Theta$ at $\Theta = 0$. The calculation can be done by identifying $A_p^\dag
A_q \Lambda_q^{\ell^{'}} B_q (\Lambda_p^{\ell} B_p)^\dag$ as
$\Phi$ and applying Theorem \ref{them2}.

In the same spirit, we can update $\Lambda_p^{\ell}$ as $(\Lambda_p \Lambda)^{\ell}$ and compute the partial
derivatives of \begin{eqnarray*} &  & P \left ( A_p (\Lambda_p \Lambda)^{\ell} B_p, \; A_q \Lambda_q^{\ell^{'}}
B_q \right )\\
& = & P \left ( \Lambda^{\ell}, \; (A_p \Lambda_p^\ell)^\dag A_q \Lambda_q^{\ell^{'}} B_q B_p^\dag \right )
\end{eqnarray*} with respect to $\Lambda$ at $\Lambda = I$ (i.e., $\lambda_m = 0, \; m = 1, \cdots, M$).  This
can be accomplished by letting
\[
\Phi = (A_p \Lambda_p^\ell)^\dag A_q \Lambda_q^{\ell^{'}} B_q B_p^\dag \]
 and invoking Theorem \ref{them2}.

Finally, because of symmetry,  we have \[ P \left ( A_p \Lambda_p^{\ell} B_p, \; A_q \Lambda_q^{\ell^{'}} B_q
\right ) = P \left ( A_q \Lambda_q^{\ell^{'}} B_q, \; A_p \Lambda_p^{\ell} B_p \right ). \]
  Hence, we can
update matrices $A_q, \; \Lambda_q$ and $B_q$ and compute the
corresponding partial derivatives by the similar method as that of
matrices $A_p, \; \Lambda_p$ and $B_p$.

In the gradient-based optimization, we used the standard steepest
gradient descent method in \cite{Zwillinger}, with some minor
modification  to adapt to parameter bounds.   In the course of
experimenting with the new design paradigm, we have observed that
it is beneficial to apply the following searching strategy.

\begin{description}

\item  \underline{STEP (a).} Find the best constellation of
diagonal structure $\{ \Lambda^\ell \; | \; 0 \leq \ell \leq
L-1\}$.  This can be done as follows. First, perform random search
to find $n$ good initial values of $\Lambda$. Second, for each
initial value of $\Lambda$, perform gradient-based optimization.
Finally,  choose the best one among the $n$ outcomes.

\item \underline{STEP (b).}  Let $\Lambda$ be found in the first
step. Find the best constellation of special structure $\{
\Lambda^\ell B_q \; | \; \ell = 0, \cdots, L-1;  \;q = 0, \cdots,
2^b -1 \}$ by employing gradient descent search over $B_q$ while
$\Lambda$ is fixed.  Here the initial value of $B_q$ can be
randomly chosen.

\item \underline{STEP (c).} Using the code found at the second
step as starting point, search $A_q, \; B_q$ and $\Lambda_q$ by
gradient descent method. Here the initial value of $A_q$ can be
randomly chosen.

\end{description}

For Steps (a)-(c) in the above strategy, we have adopted the same
choice of the step size as that of the algorithm of
\cite{Zwillinger}.

\section{Fast Decoding} \label{decoding}

Now that we have efficient constellation design tools, we focus on the all important decoding problem. In this
section, we  develop efficient algorithms for decoding our proposed new codes. Interestingly, such decoding
algorithms are also applicable to existing codes. For  ease of presentation, we first focus on the case that the
constellation has only one block (i.e., $b = 0$) and the receiver is equipped with only one antenna (i.e., $N =
1$). Subsequently, we discuss the decoding for the general cases of multiple blocks and multiple receiver
antennas (i.e., $b \geq 0$ and $N \geq 1$).

When $b = 0$, the constellation reduces to the continuous diagonal
code. The signal constellation consists of $L$ diagonal matrices
$V_\ell = \Lambda^\ell, \;\; \ell = 0, 1, \cdots, L-1$. For $N=1$,
the received signal $X_\tau \in \mathbb{C}^{M \times 1}$ is a
complex vector.  As described in \cite{Clarkson}, the ML decoding
problem can be reformulated as a problem of minimizing a Euclidean
norm as follows:
\begin{eqnarray}
&   & \widehat{z}^\mathrm{ML}_\tau \nonumber \\
 & = & \arg  \; \min_{\ell} \; || X_\tau -
V_\ell X_{\tau -1} ||_{\mathrm{F}}^2 \nonumber\\
& \approx &  \arg  \min_{\ell} \sum_{m =1}^M [(C_m \lambda_m  \ell - C_m \varphi_m) \; {\rm mod}^* C_m L]^2
\label{app}
\end{eqnarray}
where
\[
C_m = \sqrt{ \left| [X_\tau]_{m1} \; [X_{\tau-1}]_{m1} \right|}, \quad \varphi_m = \arg \left(
\frac{[X_\tau]_{m1} } { [X_{\tau-1}]_{m1} } \right) \frac{L} {2 \pi}.
\]
It has been demonstrated in \cite{Clarkson} that the approximation
in (\ref{app}) is extremely accurate. Therefore, the decoding
problem for the case $b=0, \;\; N=1$ has been transformed into the
minimization problem of finding \be \widehat{z}^\mathrm{eucl} =
\arg  \; \min_{\ell} \; \sum_{m =1}^M [(C_m \lambda_m \; \ell -
C_m \; \varphi_m) \; {\rm mod}^* C_m L]^2. \label{Euclid} \ee In
the following sub-sections we develop an efficient algorithm for
this minimization problem.

\subsection{Lattice Decoding Algorithms}

In the special case that $\lambda_m,  \; m = 1, \cdots, M$ are
integers, the continuous diagonal code reduces to the cyclic group
code. In order to decode the cyclic group code, Clarkson {\it et
al.,} \cite{Clarkson} developed an approximate solution for the
minimization problem (\ref{Euclid}). The key steps are as follows:
\begin{enumerate}
\item  Reformulate  minimization problem (\ref{Euclid}) as a
lattice closest point problem \be \arg \; \min_{y \in
\mathbb{Z}^{1 \times M} }  \; ||y G - \xi|| \label{lat} \ee where
$\xi = [\xi_1, \cdots, \xi_M]$ with $\xi_m = C_m \;  \varphi_m,
\;\; m = 1, \cdots, M$ and $G$  is a $M \times M$ generator matrix
such that
\[
[G]_{p q} = \left\{  \begin{array}{ll}
C_q \lambda_q & {\rm for} \; p = 1 \; {\rm and} \; 1 \leq q \leq M,\\
C_q L & {\rm for} \; 1 < p = q \leq M, \\
0 & {\rm else}. \end{array}\right.
\]
\item  Apply the ``LLL'' lattice algorithm \cite{Lenstra} to find
an approximate solution $\widetilde{y} = [\widetilde{y}_1, \cdots,
\widetilde{y}_M]$ for (\ref{lat}). An estimate for
$\widehat{z}^\mathrm{eucl}$ is taken as $\widetilde{y}_1 \; {\rm
mod} \; L$.
\end{enumerate}

While the ``LLL'' lattice algorithm approximately solves
(\ref{lat}), existing sphere decoder algorithms (see, e.g.,
\cite{Agrell}, \cite{Damen1, Damen3}, \cite{Fincke},
\cite{Viterbo} and the references therein) can provide an exact
solution for (\ref{lat}) and hence improve decoding accuracy. The
sphere decoder takes advantage of the lattice structure of the
received signals and proceeds as follows: (i) It searches the
closest lattice points to the received signal which are enclosed
in a sphere centered at the received signal; (ii) each time a
lattice point of a smaller norm is found, it reduces the sphere
radius accordingly and restart the search until an empty sphere is
reached. The choice of initial radius depends on the lattice
considered, as well as on the additive noise level. At the heart
of the sphere decoder algorithm is the subroutine which serves the
purposes of: (a) determining whether a sphere $|| y \; G - \xi||^2
< \gamma^2$ with fixed radius $\gamma
 >0$ is empty; (b) detecting a vector in it otherwise.
A Cholesky factorization is performed to find an upper triangular matrix $D$ so that $D^\intercal D = G
G^\intercal$, from which the boundary conditions of the sphere can be derived as \be z_k - \sqrt{
\frac{\vartheta_k}{t_{kk}}} - \varpi_k < y_k <   \sqrt{ \frac{\vartheta_k}{t_{kk}}} - \varpi_k + z_k, \; k = M,
M-1, \cdots,  1 \label{bbb} \ee where
\[
[z_1, \cdots, z_M] = \xi G^{-1}, \; \varpi_k = \sum_{j = k + 1}^M t_{k j} (y_j - z_j), \; k = 1, \cdots, M
\]
and
\[
\vartheta_M = \gamma^2, \;\;\; \vartheta_{k-1} = \vartheta_k -
t_{kk} \; (y_k- z_k + \varpi_k)^2, \;\;\; k = 2, \cdots, M
\]
with $t_{kk} = [D]_{kk}^2, \;\; k=1,\cdots, M$ and $t_{kj} = \frac{ [D]_{kj} } { [D]_{kk}}, \;\; 1 \leq k < j
\leq M$ (see, e.g., \cite{Fincke, Viterbo} for details). Clearly, the boundary of $y_k$ depends on values of
$y_{j}, \;\; j = k+1, \cdots, M$. If the set of feasible values of $y_M$, denoted by $\mathcal{ I}_M$, is not
empty, then for each member $y_M$ of $\mathcal{ I}_M$ the values of other coordinates needed to be evaluated in
the sphere decoder algorithm can be represented as the nodes of a tree starting from $y_M$. The following Figure
\ref{fig000} depicts this tree structure.  In the tree, the children nodes are generated from the parent nodes
in accordance with the boundary equation (\ref{bbb}). A path of length $M$ (i.e., consisting of $M$ nodes)
corresponds to a vector located in the sphere. When $\mathcal{ I}_M$ has multiple members, the task of the core
subroutine is to search among the multiple trees to determine whether there is a path of length $M$ and identify
one if there exists. It should be noted that, in sphere decoding, {\it most of the computational efforts are
devoted to the evaluation of paths of length less than $M$}.

\begin{figure}
\centering
\includegraphics[width=3.3in]{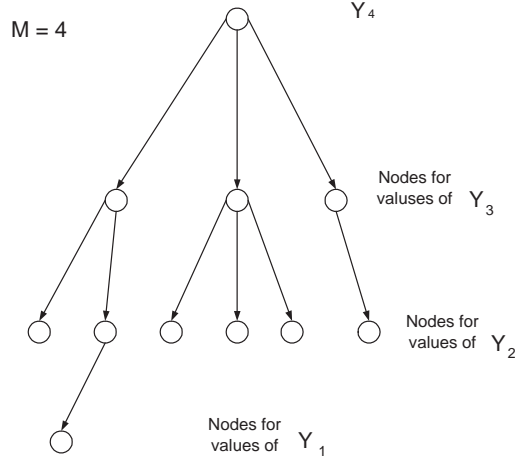}
\caption{A tree representation of values of coordinates to be
investigated for a fixed $y_M$.} \label{fig000}
\end{figure}

\subsection{Removing the Curse of Dimensionality}

In the general case that $\lambda_m,  \; m = 1, \cdots, M$ are
continuous parameters, the minimization problem in (\ref{Euclid})
lacks the lattice structure.  Hence, existing sphere decoder
algorithms and ``LLL'' lattice algorithm are not applicable.
Moreover,  even for the special case of cyclic group code, the
lattice decoding algorithms described in the last subsection aim
to solve a closest point problem of dimension $M$ (the dimension
will be expanded to $MN$ when using $N$ receiver antennas). The
computational complexity may be too high when the number of
transmitter antennas $M$ (or the number of receiver antennas $N$)
is large. Therefore, it is crucial to reduce the dimension of the
underlying closest point problem by further exploiting the
diagonal structure of signal constellation. We achieve the
reduction and improve efficiency with a new decoding algorithm,
applicable to the general case that $\lambda_m,  \; m = 1, \cdots,
M$ are continuous parameters. As a critical step to reduce
decoding complexity, we will show next that the dimension of the
related closest point problem can be reduced to one.

\bigskip

\begin{theorem}  \label{reduce}
Define
\[
\begin{array}{ll}
{\mathcal S} & =  \; \{ (y_1, \cdots, y_M) \; | \: y_1 \in \mathbb{Z}\\
& \qquad  {\rm and} \;
  - \frac{L}{2} + \varphi_1 \leq y_1 < \frac{L}{2} + \varphi_1; \\
  & \qquad y_m = \left  \lceil \frac{\varphi_m } { L }  -
( \frac{y_1}{L} - \lfloor \frac{y_1}{L} \rfloor  ) \lambda_m - \frac{1}{2} \right \rceil - \lfloor \frac{y_1}{L}
\rfloor \lambda_m\\
& \qquad {\rm for} \;\; m = 2, \cdots, M \}. \end{array}
\]
Suppose that there exists an unique $\widehat{\ell} \in \{0, 1, \cdots, L-1\}$ such that \begin{eqnarray*}
&   & \sum_{m =1}^M [(C_m \lambda_m \; \widehat{\ell} - C_m \varphi_m) \; {\rm mod}^* C_m L]^2\\
& = & \min_{\ell} \; \sum_{m =1}^M [(C_m \lambda_m \; \ell - C_m \varphi_m) \; {\rm mod}^* C_m L]^2.
\end{eqnarray*}
 Then
\begin{eqnarray*}
\widehat{\ell} & = & \arg  \; \min_{\ell} \; \sum_{m =1}^M [(C_m \lambda_m \; \ell - C_m \varphi_m) \; {\rm
mod}^* C_m L]^2\\
& = & \widehat{y}_1 - \left \lfloor \frac{ \widehat{y}_1 } { L } \right \rfloor \; L
\end{eqnarray*} where $\widehat{y}_1$ is first entry of
\[
\widehat{y} = [\widehat{y}_1, \cdots, \widehat{y}_M] = \arg \; \min_{y \in \mathcal{ S} } || y G - \xi||^2.
\]
\end{theorem}

\bigskip

See Appendix \ref{app3} for a proof.

It can be seen from Theorem \ref{reduce} that $y_q$, for $q = 2, \cdots, M$, is uniquely determined by $y_1$.
Hence, finding \be \arg \; \min_{y \in \mathcal{ S} } || y G - \xi||^2 \label{closest} \ee is essentially a
one-dimensional closest point problem.

Next, by exploiting the special structure of the constellation, we derive extremely simple boundary conditions
for the sphere $\left\{ y \in \mathcal{ S} \; \left| \;  || y G - \xi||^2 < \gamma^2 \right. \right\}$.

\bigskip

\begin{theorem}  \label{th_bbb}
Let $y = [y_1, \cdots, y_M]  \in \mathcal{ S}$. Define
\[
\mu_1 = [ C_1 \left( y_1 - \varphi_1 \right) ]^2
\]
and
\[
\mu_m = \mu_{m-1} + [C_m \left( L y_m + \lambda_m y_1 - \varphi_m \right)]^2 \] for $m = 2, \cdots, M$.
Then $
|| y G - \xi||^2 < \gamma^2$ if and only if $y_1$ is an integer satisfying \be \varphi_1 - \frac{\gamma}{C_1} <
y_1 < \frac{\gamma}{C_1} + \varphi_1, \;\;\;\; - \frac{L}{2} + \varphi_1 \leq y_1 < \frac{L}{2} + \varphi_1
\label{boundary0} \ee and \be \mu_m < \gamma^2\;\; {\rm for} \;\; m = 2, \cdots,M. \label{boundary} \ee
\end{theorem}

\bigskip

See Appendix \ref{app4} for a proof.

It can be seen that the conditions  in (\ref{boundary0}) determine an interval $\mathcal{ I}_1$ of feasible
values for $y_1$. For each value of $y_1 \in \mathcal{ I}_1$, we only need  to evaluate the simple conditions in
(\ref{boundary}). This is in sharp contrast to the search over a tree structure described above in the context
of sphere decoder.

It should be noted that the sphere decoding algorithm is
originally devised to find closest lattice points. In general, our
decoding problem is not a problem of searching closest lattice
points. However, we can still use the sphere decoding algorithm
because the enumeration of interior points of a sphere can be
efficiently done as the case of a lattice problem. Moreover,
Theorem \ref{th_bbb} indicates that the ``sphere'' can actually be
reduced to an ``interval'' of one dimension.

\subsection{Simplified Sphere Decoder}

Using the reduction of dimensionality described in the last
subsection, we now develop a new decoding algorithm which also
applies to continuous diagonal codes, FPF codes $G_{m,r}$,
non-group codes $S_{m,s}$ and products of groups. To further
enhance efficiency, we adopt the ``zigzag'' searching strategy
originated in \cite{Schnorr} and the idea proposed in
\cite{Damen3} for avoiding repeated computations.

Obviously, the search for $y_1$ can significantly affect the efficiency. Let $\widehat{y}$ denote the vector
corresponding to the transmitted signal. Intuitively, for moderate and high SNR, it is more likely for the
received signal $\xi$ to be closer to $\widehat{y} \; G$. Since $|y_1 - \varphi_1| < ||y G - \xi||$, we should
first investigate $y_1$ which is closer to $\varphi_1$ for a better chance of detecting $\widehat{y}$.
Therefore, we shall investigate $y_1$ in the following sequence, \be \lfloor \varphi_1 \rceil + (-1)^k\left
\lfloor \frac{k}{2} \right \rfloor, \;\; k = 0, 1, 2, \cdots. \label{rep} \ee That is, the investigation is
started from $\lfloor \varphi_1 \rceil$ and proceeded in a ``zigzag'' order in the outward directions (see,
e.g., \cite{Agrell}, \cite{Schnorr}).   Note that  condition (\ref{boundary0}) implies
\[
- \left \lceil \min \left( \frac{\gamma}{C_1}, \; \frac{L}{2}
\right) \right \rceil \leq y_1 - \lfloor \varphi_1 \rceil \leq
\left \lceil \min \left( \frac{\gamma}{C_1}, \; \frac{L}{2}
\right) \right \rceil.
\]
Hence, it suffices to investigate
\[
y_1 = \lfloor \varphi_1 \rceil + (-1)^k\left \lfloor \frac{k}{2}
\right \rfloor, \;\; k = 0, 1, \cdots, 2 \left \lceil  \min \left(
\frac{\gamma}{C_1}, \; \frac{L}{2} \right) \right \rceil.
\]

It is also important to avoid repeated investigation of $y_1$.
When a value of $y_1$ is found to satisfy
condition~(\ref{boundary}), the radius $\gamma$ is reduced as
$\sqrt{\mu_M}$ and the interval confining $y_1$ is consequently
shrunk. In this way, the range of $y_1$ needed to be investigated
is squeezed from outside. To improve efficiency, we use the idea
of \cite{Damen3} to ensure that the range of $y_1$ needed to be
investigated is also squeezed from inside. The idea is based on
the following observation:

{\it For a given radius $\gamma$, if a value of $y_1$ violates the
boundary conditions, then the same value of $y_1$ also violates
the corresponding boundary conditions after $\gamma$ is reduced.}

Therefore, we can keep a record for the values of $y_1$ which have
been investigated in order to avoid repeated computation.  For
this purpose, the index variable $k$ in (\ref{rep}) can be used as
an indicator for the range of values investigated.

In summary, the decoding algorithm is presented as follows.

\begin{description}

\item \underline{STEP 1.}  Input $\gamma \leftarrow
\gamma_\mathrm{init}$ where initial radius $\gamma_\mathrm{init}$
is chosen based on noise level. Let $k \leftarrow 0$ and
$\widetilde{y_1} \leftarrow \lfloor \varphi_1 \rceil$.

\item \underline{STEP 2.}  Let $k_\mathrm{max} \leftarrow 2 \left
\lceil \min \left( \frac{\gamma}{C_1}, \; \frac{L}{2} \right)
\right \rceil$.

\item  \underline{STEP 3.}  If $k \leq k_\mathrm{max}$, let $y_1
\leftarrow \lfloor \varphi_1 \rceil + (-1)^k \left \lfloor
\frac{k}{2} \right \rfloor$ and $k \leftarrow k + 1$. Otherwise,
let $\gamma \leftarrow \frac{4}{3} \gamma, \;\;\; k \leftarrow 0$
and go to STEP 2.

\item  \underline{STEP 4.}  If condition~(\ref{boundary}) is
violated, go to STEP 3. Otherwise, let $\widetilde{y_1} \leftarrow
y_1, \;\;\; \gamma \leftarrow \sqrt{\mu_M}$.

\item \underline{STEP 5.}  Using $\gamma, \; \widetilde{y_1}$ and
$k$ as input, call subroutine CLOSEST-POINT to find
$\widehat{y_1}$.  Then $\widehat{z}^\mathrm{eucl}$ is calculated
as $\widehat{z}^\mathrm{eucl} = \widehat{y}_1 - \left \lfloor
\frac{ \widehat{y}_1 } { L } \right \rfloor  L$.  Return
$\widehat{z}^\mathrm{eucl}$ as the estimate and stop.

\end{description}

The subroutine CLOSEST-POINT is presented as follows.

\begin{description}
\item  \underline{Function: CLOSEST-POINT}

\item \underline{STEP 1.}  Input $\gamma, \; \widetilde{y_1}$ and
$k$.  Let $\widehat{y_1} \leftarrow \widetilde{y_1}$.

\item \underline{STEP 2.}  Let $k_\mathrm{max} \leftarrow 2 \left
\lceil  \min \left( \frac{\gamma}{C_1}, \; \frac{L}{2} \right)
\right \rceil$.

\item  \underline{STEP 3.}  If $k \leq k_\mathrm{max}$, let $y_1
\leftarrow \lfloor \varphi_1 \rceil + (-1)^k \left \lfloor
\frac{k}{2} \right \rfloor$ and $k \leftarrow k + 1$. Otherwise,
go to STEP 5.

\item  \underline{STEP 4.}  If condition~(\ref{boundary}) is
satisfied, then let $\widehat{y_1} \leftarrow y_1, \;\; \gamma
\leftarrow \sqrt{\mu_M}$ and go to STEP 2.  Otherwise, go to STEP
3.

\item \underline{STEP 5.}  Return $\widehat{y_1}$ and stop.

\end{description}

It can be seen from STEP 3 that the index $k$ has served the
purpose of avoiding repeated investigations.  Once a nonempty
sphere is detected, no value of $y_1$ is investigated more than
once among the subsequent smaller spheres.

It should be also noted that, compared to conventional sphere
decoder algorithms, many computationally expensive steps have been
avoided in our algorithm. For examples,  the Cholesky
factorization of $G G^\intercal$ and the computation of $\xi
G^{-1}$ are not needed in our algorithms.

\subsection{Sphere Decoding -- The General Case}

We now discuss the decoding problem for the general case of multiple receiver antennas and multiple block
constellation (i.e., $N \geq 1, \; b \geq 0$).  Since the ML-decoding is computationally difficult, our goal is
to develop a sub-optimal decoding method with low decoding complexity. Note that the ML decoding finds \be \arg
\; \min_{\ell, q} ||X_\tau - A_q \Lambda_q^\ell B_q X_{\tau -1} ||_{\mathrm{F}}^2, \label{mut_bb} \ee which can
be done by obtaining \be \arg \; \min_{\ell} ||X_\tau - A_q \Lambda_q^\ell B_q X_{\tau -1} ||_{\mathrm{F}}^2
\label{route} \ee for $q = 0, 1, \cdots, 2^b -1$ and  seeking the tuple $(\widehat{q}, \widehat{\ell})$
minimizing the Frobenius norm. This method is of sequential nature and has been used in \cite{Shokrollahi} for
decoding FPF code $G_{m,r}$, non-group code $S_{m,s}$ and products of groups with  the ``LLL'' lattice algorithm
sequentially applied to solve (\ref{route}).

In the general case the underlying closest point problem is of
dimension $MN$ and sequential approaches are inefficient. We use
the simplified sphere decoder algorithm developed in the previous
subsection and transform the decoding problem into $2^b$
one-dimensional closest point problems that can be solved in
parallel.

Since the Frobenius norm of a matrix is invariant under unitary
transformations, we have
\begin{eqnarray*}
||X_\tau - A_q \Lambda_q^\ell B_q X_{\tau -1} ||_{\mathrm{F}}^2 &
= & ||
A_q^\dag (X_\tau - A_q \Lambda_q^\ell B_q X_{\tau -1}) ||_{\mathrm{F}}^2\\
& = & ||A_q^\dag X_\tau - \Lambda_q^\ell B_q X_{\tau -1}
||_{\mathrm{F}}^2.
\end{eqnarray*}
By a similar method as that of \cite{Clarkson}, we can show that
\begin{eqnarray}
&   & ||A_q^\dag X_\tau - \Lambda_q^\ell B_q X_{\tau -1}
||_{\mathrm{F}}^2
\nonumber\\
& = & \sum_{n=1}^N \sum_{m = 1}^M \left| [A_q^\dag X_\tau]_{mn} -
e^{i 2 \pi \lambda_{q,m} \ell \slash L} \; [B_q  X_{\tau -1}]_{mn}
\right|^2
\nonumber\\
& = & ||X_\tau ||_{\mathrm{F}}^2 + ||X_{\tau -1}||_{\mathrm{F}}^2\nonumber\\
&   & - 2 \sum_{n=1}^N \sum_{m=1}^M  C_{m,n}^2 \cos ([ ( \lambda_{q,m} \; \ell -
\varphi_{m,n}) \; {\rm mod}^* L ]2 \pi \slash L) \nonumber\\
& \approx & ||X_\tau ||_{\mathrm{F}}^2 + ||X_{\tau -1}||_{\mathrm{F}}^2 - 2 \sum_{n=1}^N \sum_{m=1}^M
C_{m,n}^2\nonumber\\
&   &  + \sum_{n=1}^N \sum_{m=1}^M  C_{m,n}^2 ([ ( \lambda_{q,m} \; \ell -
\varphi_{m,n}) \; {\rm mod}^* L ]2 \pi \slash L)^2 \nonumber\\
& = & \frac{4 \pi^2}{L^2}  \Delta_q + \frac{4 \pi^2}{L^2} \times
\nonumber\\
&   & \sum_{n=1}^N \sum_{m=1}^M [ (C_{m,n} \lambda_{q,m} \ell -
C_{m,n} \varphi_{m,n}) \; {\rm mod}^* C_{m,n} L ]^2 \nonumber\\
\label{mult_ap}
\end{eqnarray}
where
\[
C_{m,n} = \sqrt{\left| [A_q^\dag X_\tau]_{mn} \; [B_q  X_{\tau -1}]_{mn} \right| },
\]
\[
\varphi_{m,n} = \arg \left( \frac{[A_q^\dag X_\tau]_{mn}} {[B_q  X_{\tau -1}]_{mn}} \right)  \frac{L } {2 \pi},
\]
and
\[
\Delta_q = \frac{ L^2 ||{\rm abs} (A_q^\dag X_\tau) - {\rm abs} (
B_q X_{\tau -1} )||_{\mathrm{F}}^2  } {4 \pi^2}.
\]
Define a $MN \times MN$ matrix $G^q$ such that
\[
[G^q]_{k j} = \left\{  \begin{array}{l} C_{ (j - \lfloor \frac{j-1}{M} \rfloor M, \; \lfloor \frac{j-1}{M}
\rfloor + 1 )} \; \lambda_{(q, \; j - \lfloor \frac{j-1}{M} \rfloor M)} \\
 {\rm for} \; k =
1 \; {\rm and} \; 1 \leq j \leq MN;\\
\\
C_{ (j - \lfloor \frac{j-1}{M} \rfloor M, \; \lfloor \frac{j-1}{M}
\rfloor +
1 )}  \; L\\
 {\rm for} \; 1 < k = j \leq MN; \\
 \\
0 \; {\rm else.} \end{array}\right.
\]
Define a row vector $\xi^q = [\xi_1, \cdots, \xi_{MN}]$ such that
\[
\xi_k = C_{ (k - \lfloor \frac{k-1}{M} \rfloor M, \; \lfloor \frac{k-1}{M} \rfloor + 1 )  } \; \varphi_{ (k -
\lfloor \frac{k-1}{M} \rfloor M, \; \lfloor \frac{k-1}{M} \rfloor + 1 ) } \] for $k = 1,\cdots, MN$.  Define
$\psi_k = \varphi_{ (k - \lfloor \frac{k-1}{M} \rfloor M, \; \lfloor \frac{k-1}{M} \rfloor + 1 ) }$ and $\beta_k
= \lambda_{ (q, \; \lfloor \frac{k-1}{M} \rfloor + 1 ) }$ for $k = 1,\cdots, MN$. Define
\[
\begin{array}{ll}
\mathcal{ S}^q & =  \; \{ (y_1, \cdots, y_{MN}) \; | \: y_1 \in \mathbb{Z}\\
&  \qquad {\rm and} \;
  - \frac{L}{2} + \psi_1 \leq y_1 < \frac{L}{2} + \psi_1; \\
  & \qquad y_k = \left  \lceil \frac{\psi_k } { L }  -
( \frac{y_1}{L} - \lfloor \frac{y_1}{L} \rfloor  ) \beta_k - \frac{1}{2} \right \rceil - \lfloor \frac{y_1}{L}
\rfloor \beta_k\\
& \qquad {\rm for} \;\; k = 2, \cdots, MN \}. \end{array}
\]
Then, by Theorem \ref{reduce}, we have \begin{eqnarray} &   & \min_{\ell} \sum_{n=1}^N \sum_{m=1}^M [ (C_{m,n}
\lambda_{q,m} \ell - C_{m,n} \varphi_{m,n}) \; {\rm mod}^* C_{m,n} L ]^2 \nonumber\\
& = & \min_{y \in \mathcal{ S}^q}  || y G ^q - \xi^q ||^2. \label{mut2} \end{eqnarray} It follows from
(\ref{mult_ap}) and (\ref{mut2}) that
\begin{eqnarray*}
&   & ||A_q^\dag X_\tau - \Lambda_q^\ell B_q X_{\tau -1} ||_{\mathrm{F}}^2\\
& \approx & \frac{4 \pi^2}{L^2} \; \min_{y \in \mathcal{ S}^q} \; \left( || y G ^q - \xi^q ||^2 + \Delta_q
\right),
\end{eqnarray*}
leading to
\begin{eqnarray*}
&    & \min_{\ell, q} ||X_\tau - A_q \Lambda_q^\ell B_q X_{\tau -1} ||_{\mathrm{F}}^2\\
& \approx & \frac{4 \pi^2}{L^2} \; \min_{q} \; \min_{y \in \mathcal{ S}^q} \; \left(  || y G ^q - \xi^q ||^2 +
\Delta_q \right).
\end{eqnarray*}
Hence, by Theorem \ref{reduce}, the maximum likelihood decoder can be well approximated by
\[
(\widehat{q}, \widehat{\ell}) = \left(\widehat{q}, \;
\widehat{y}_1 - \left \lfloor \frac{\widehat{y}_1}{L} \right
\rfloor \right)
\]
where $\widehat{y}_1$ is the first entry of $\widehat{y} =
[\widehat{y}_1,\cdots, \widehat{y}_{MN}]$ such that
\[
(\widehat{q},\widehat{y}) =  \arg \; \min_{q} \; \min_{y \in \mathcal{ S}^q} \; \left( || y G ^q - \xi^q ||^2 +
\Delta_q \right).
\]
The above analysis shows that the efficiency of the decoding
problem (\ref{mut_bb}) can be enhanced by sequentially applying
the simplified sphere decoder algorithms developed in the last
subsection. In the next sub-section we shall improve the
efficiency even further by developing a parallel search strategy.

\subsection{Parallel Sphere Decoding}

The sequential sphere decoder algorithm introduced in the last
subsection involves $2^b$ independent sphere decoding processes.
When the constellation consists of many blocks (i.e., large
$2^b$), the sequential decoding may be too time consuming, but
given the independence of each sphere decoding all searches can be
executed in parallel. Specifically, since it has been shown in the
previous subsection that the ML decoding problem (\ref{mut_bb})
can be reformulated as the sub-optimal decoding problem
\[
\arg \; \min_{q = 0, 1, \cdots, 2^b - 1} \; \min_{y \in \mathcal{ S}^q}  \; \left(  || y G^q - \xi^q||^2 +
\Delta_q \right),
\]
we can apply in parallel the simplified sphere decoder algorithm
to investigate the following $2^b$ spheres:
\[
\left\{ y \in \mathcal{ S}_q \; \left| \;  || y G^q - \xi^q||^2 < \gamma_q^2 \right. \right\},  \;\;\;\; q = 0,
1, \cdots, 2^b - 1
\]
where \be \gamma_q^2 = \gamma^2 - \Delta_q, \;\;\;\; q = 0, 1, \cdots, 2^b - 1 \label{shrink} \ee with parameter
$\gamma$ controlling the sizes of all spheres.  The choice of the initial value of $\gamma$ is similar to
choosing the initial radius of conventional sphere decoder. For a fixed value of $\gamma$, the $2^b$ spheres
respectively determine $2^b$ sets of feasible $y_1$ values based on (\ref{boundary0}). Let the set for the
$q$-th sphere be denoted as $\mathcal{ I}^q$. We investigate the $y_1$ values of these sets in a round robin
order.  That is, the sets are visited in the following sequence
\[
\mathcal{ I}^0, \mathcal{ I}^1, \cdots, \mathcal{ I}^{2^b-1}; \; \mathcal{ I}^0, \mathcal{ I}^1, \cdots,
\mathcal{ I}^{2^b-1}; \; \cdots \cdots.
\]
Of course, any value of $y_1$ will be eliminated from its corresponding set after evaluation. Once a value of
$y_1$ from set $\mathcal{ I}^q$ is found to guarantee (\ref{boundary0}) and (\ref{boundary}), $\gamma_q$ is
reduced as $\sqrt{\mu_M}$. Subsequently, $\gamma$ is reduced as $\sqrt{\gamma_q^2 + \Delta_q}$ and the radius of
other spheres are decreased accordingly by (\ref{shrink}). When no value of $y_1$ from any set satisfies
(\ref{boundary0}) and (\ref{boundary}), $\gamma$ will be increased and consequently all the spheres are enlarged
based on (\ref{shrink}). All the spheres keep enlarging before detecting a value of $y_1$ guaranteeing
(\ref{boundary0}) and (\ref{boundary}). Once the value of $y_1$ is found, all spheres begin to shrink. The
shrinking process is very quick due to the parallel mechanism. This process is terminated when all these sets
become empty.  The solution of decoding problem (\ref{mut_bb}) is given as the tuple
\[
\left( \widehat{q}, \;\; \widehat{y}_1 - \left \lfloor \frac{
\widehat{y}_1 } { L } \right \rfloor L  \right)
\]
where $\widehat{y}_1 \in \mathcal{ I}^{\widehat{q}}$ and $\widehat{y}_1$ is last value found to guarantee
(\ref{boundary0}) and (\ref{boundary}). It should be noted that, in this decoding process, {\it only one CPU
processor is needed}.

\section{Illustrative Examples} \label{examples}

In this paper, we only design constellations for the special structure that $\Lambda_q = \Lambda, \;\; A_q = I$
for $q = 0, \cdots, 2^b -1$.  The computational effort has been significantly reduced by applying Theorem
\ref{th_diag}. Better codes (with lower bit error rate but equivalent decoding complexity) can be obtained if we
allow general $A_q$ and $\Lambda_q$. However, the searching time will be substantially increased if the
constellation size is large. Even in this limited case, the new design paradigm generates unitary space-time
constellations which significantly outperform existing ones.  In the following, we show the simulation results
of our codes as compared to existing codes. In comparison of the bit error rate performance, we have used the
Gray code bit mapping for the orthogonal design and CD codes, and the binary-to-decimal conversion mapping for
our codes, cyclic group codes, FPF codes and product of groups. The data of our unitary space-time codes are
reported in Appendix \ref{app5}. The details of orthogonal designs we used in our simulation is provided in
Appendix \ref{app6}.

For the case of two transmit antennas and one receiver antenna,
our computational experience indicates that it is hard to achieve
significant performance improvement upon the orthogonal design
proposed in \cite{Tarokh3}.  By using nonconstant modulus
constellations, the performance of \cite{Hwang} further improves
upon that of \cite{Tarokh3} at the price of the complexity of
estimating the channel power and signal power.  However, when the
number of transmit antennas is more than two or the number of
receiver antennas is more than one, the differential detection
scheme based on orthogonal designs subjects to significant
performance loss.

\begin{figure*}
\centering
\includegraphics[width=5.0in]{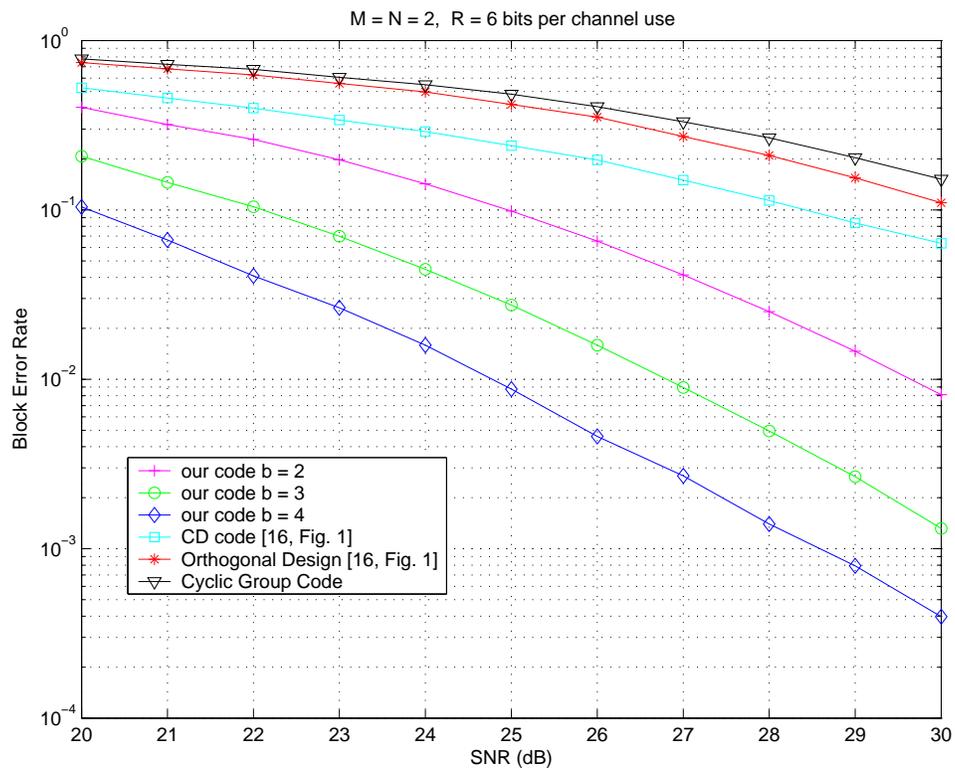}
\caption{Performance simulations of constellations} \label{fig18}
\end{figure*}

\begin{figure*} \centering
\includegraphics[width=5.0in]{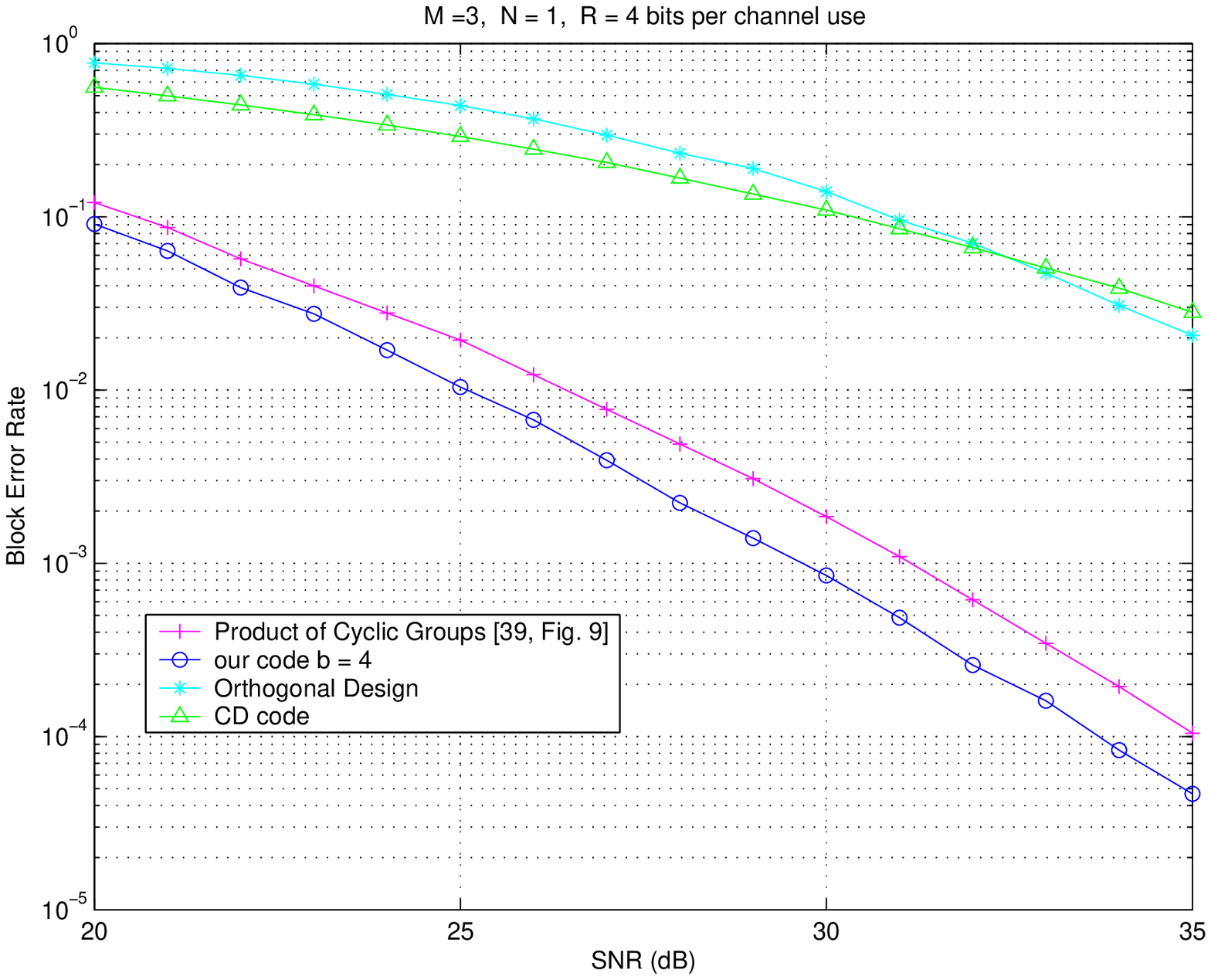}
\caption{Performance simulations of constellations} \label{fig28}
\end{figure*}

We compared the performance of our code with the differential detection schemes using orthogonal designs in
Figures 3-7.  In general, our codes significantly outperform orthogonal designs at the price of relatively
higher decoding complexity.  It can be seen from Figure \ref{fig18} that, with spectral efficiency $R =6$ bits
per channel use, our code (with block number $16$, i.e., $b=4$) improves upon orthogonal design over 10 dB at
block error rate $10^{-1}$ when using two transmitter antennas and two receiver antennas.  It is shown in Figure
\ref{fig28} that, with spectral efficiency $R =4$ bits per channel use, our code improves upon orthogonal design
about $11$ dB at block error rate $2 \times 10^{-2}$ when using $3$ transmitter antennas and one receiver
antenna.  It can be seen from Figure \ref{fig48} that, with spectral efficiency $R =3$ bits per channel use, our
code improves upon orthogonal design about 6 dB at bit error rate $10^{-3}$ when using $4$ transmitter antennas
and $2$ receiver antennas.  These examples demonstrate that orthogonal designs suffer from substantial
performance penalty. Such penalty becomes more sever when using multiple receiver antennas, or using more than
two transmit antennas, or operating at high spectral efficiency.

We compared the performance of our codes with Caley differential codes in Figures 3-7. Figure \ref{fig18} shows
that, with spectral efficiency $R =6$ bits per channel use, our code (with block number $16$, i.e., $b=4$)
improves upon Caley differential code (reported in page 1495 of \cite{Hassibi1}), about $9$ dB at block error
rate $6 \times 10^{-2}$ when using two transmitter antennas and two receiver antennas. The improvements of our
codes with block number $4$ and $8$ are respectively $4$ dB and $7$ dB at block error rate $6 \times 10^{-2}$.
The data of CD codes we used in simulation for Figures 4-7 is not available in the literature. We followed the
design method proposed in \cite{Hassibi1} to search the corresponding CD codes.  As described in
\cite{Hassibi1}, the performance metric used in the optimization is the average logarithm determinant
$\xi(\mathcal{V})$. The number of data streams $Q$ should be chosen as large as possible under constraint (30)
of \cite{Hassibi1}.  For a given spectral efficiency $R$, once $Q$ is fixed, the set $\mathcal{A}_r$ for $\{
\alpha_q \}$ is determined and is provided in \cite{Hassibi1}.  We obtained CD codes via extensive
gradient-based optimization.  The values of tuple $(Q, \xi)$ for the CD codes corresponding to Figures 4-7 are,
respectively,  $(4, 0.2610), \; (8, 0.5832), \; (12, 0.3619)$ and $(12, 0.5401)$.  As can be seen from Figures
4-7, the performance of CD codes is not comparable with that of our codes. However, we can see that CD codes are
generally better than cyclic group codes and orthogonal designs in terms of bit (or block) error rate
performance.

In Figures 4-6, we compared the performance of our proposed codes
with the FPF codes proposed in \cite{Shokrollahi}.  It is seen
from Figure \ref{fig28} that, with spectral efficiency $R =4$ bits
per channel use, our code (with block number $16$, i.e., $b=4$)
improves upon the product of cyclic groups (see Table IV of
\cite{Shokrollahi}) about $2$ dB at block error rate $2 \times
10^{-3}$ when using $3$ transmitter antennas and one receiver
antenna.  In Figure 5, our code sightly outperforms the product of
groups code. However, our code has a lower decoding complexity
since our code involves only $4$ branches of sphere decoding,
while the product of groups code involves $17$ branches of sphere
decoding.  In this case, the $T$ matrix is not available from
\cite{Shokrollahi}.  We used the same diagonal elements, $u = [1\;
3\; 4\; 11]$, as that of \cite{Shokrollahi}. We searched the best
$T$ matrix based on the conventional criterion of diversity
product maximization.  We obtained a $T$ matrix so that the
constellation has diversity product $0.3118$, which is greater
than the previously known value, $0.3105$, reported in
\cite{Shokrollahi}.  In Figure 6, our code significantly
outperforms the product of groups code.  Our code with $b=4$
improves upon the product of groups code about $3$ dB at bit error
rate $10^{-4}$.  Moreover, our code has a lower decoding
complexity since our code involves only $16$ branches of sphere
decoding, while the product of groups code involves $65$ branches
of sphere decoding.  In this case, we used the same diagonal
elements, $u = [1, \; 14, \;    21, \; 34]$,  as that of
\cite{Shokrollahi}. We searched the best $T$ matrix based on the
conventional criterion of diversity product maximization. We
obtained a $T$ matrix so that the constellation has diversity
product $0.1563$, which is greater than the previously known
value, $0.1539$, reported in \cite{Shokrollahi}.

\begin{figure*}
\centering
\includegraphics[width=5.0in]{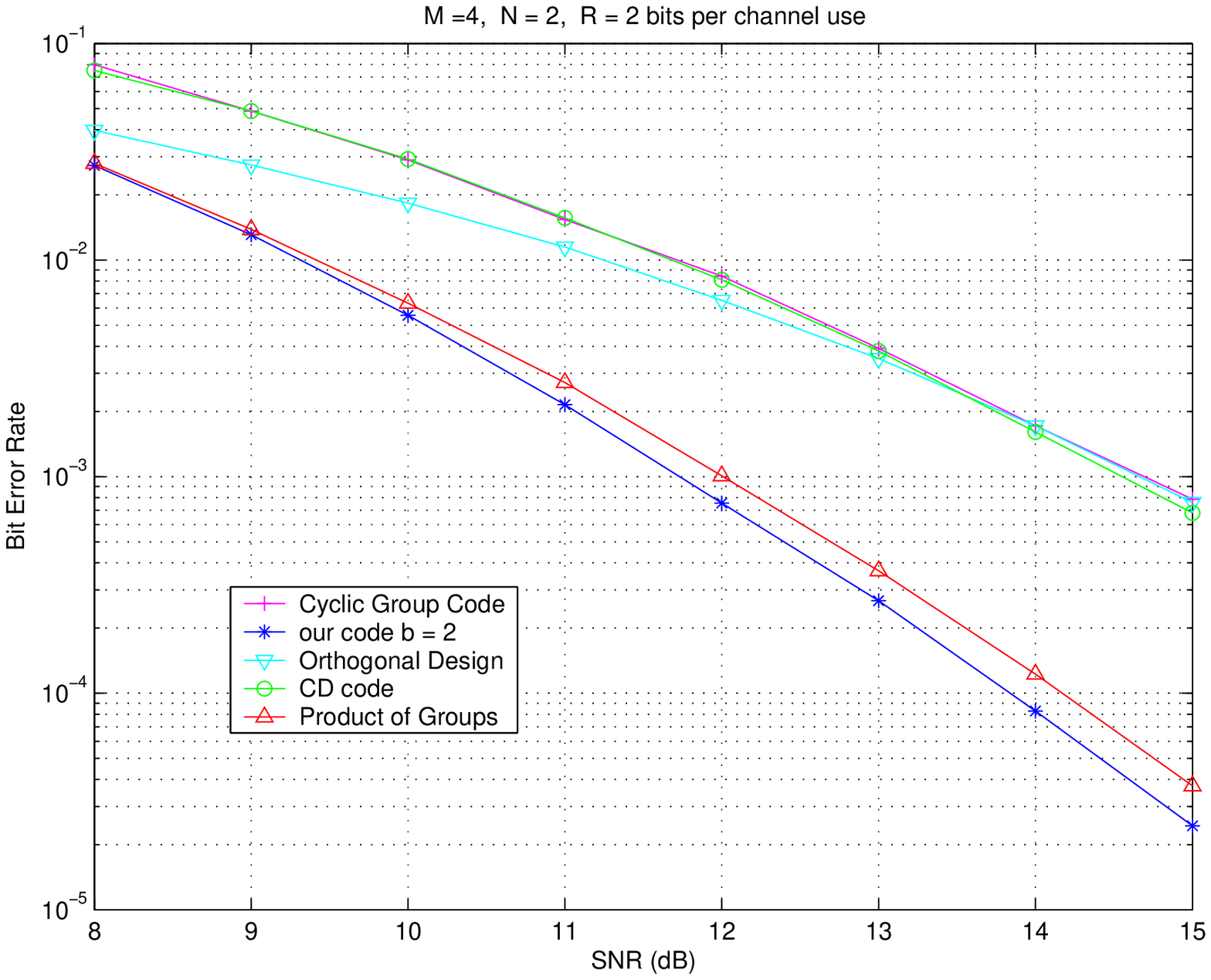}
\caption{Performance simulations of constellations} \label{fig38}
\end{figure*}

\begin{figure*}
\centering
\includegraphics[width=5.0in]{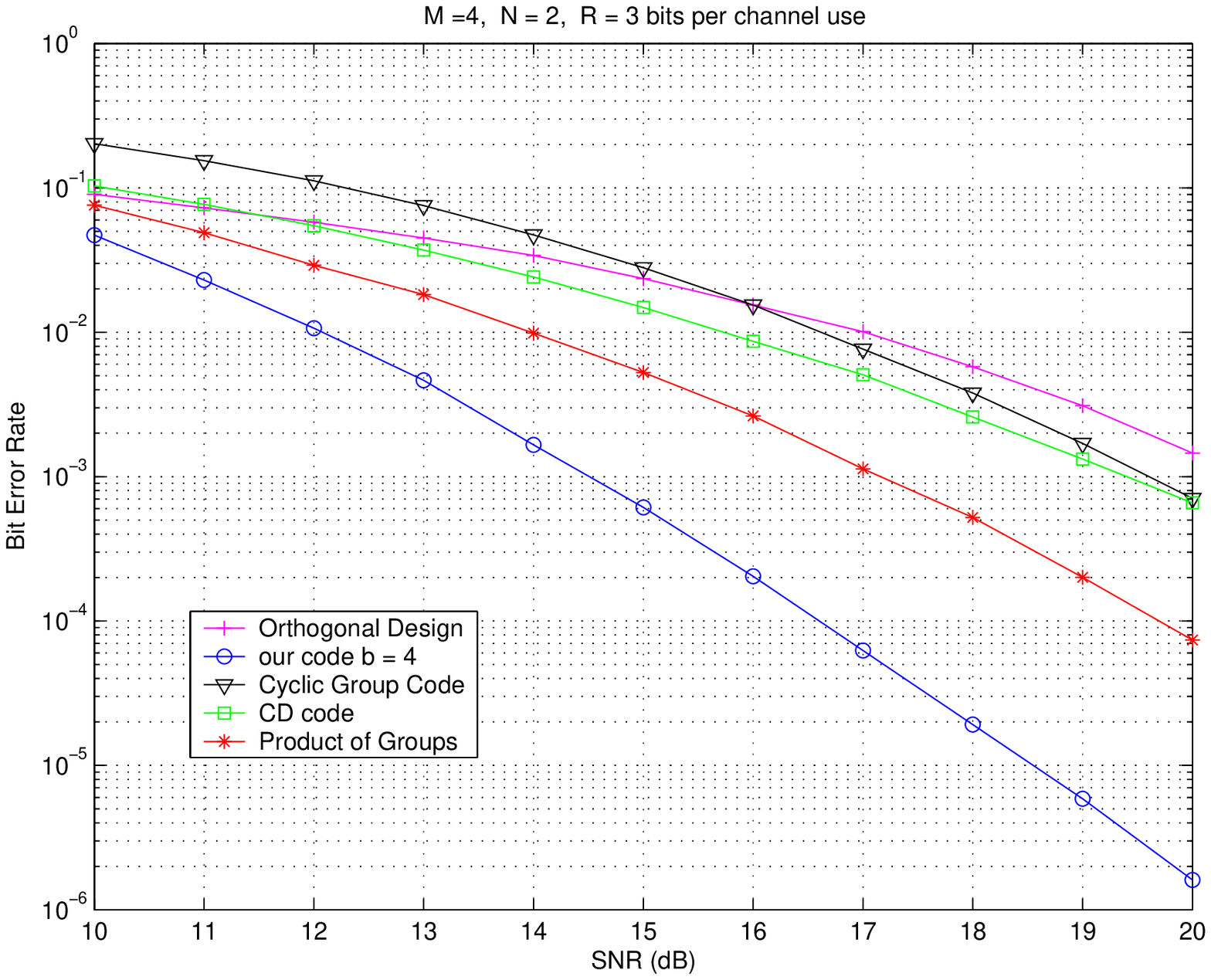}
\caption{Performance simulations of constellations} \label{fig48}
\end{figure*}

We compared the performance of our proposed code with cyclic group codes in Figures $3, \; 5$ and $6$.  It is
demonstrated that our codes significantly outperform cyclic group codes.  For example, Figure \ref{fig38} shows
that, with spectral efficiency $R =2$ bits per channel use, our code (with block number $4$, i.e., $b=2$)
improves upon the best previously known cyclic group code $u = [1 \; 25 \; 97 \; 107]$ (see Table I of
\cite{Hochwald2}) about $3$ dB at bit error rate $10^{-3}$ when using $4$ transmitter antennas and $2$ receiver
antennas.  The cyclic group code corresponding to Figure 3 is $u = [1,\; 1731]$ of diversity product $0.0265$.
The cyclic group code corresponding to Figure 6 is $u = [1, \; 301, \; 1561, \; 1829]$ of diversity product
$0.1035$. We obtained these two cyclic group codes based on the conventional criterion of diversity product
maximization.

\begin{figure*}
\centering
\includegraphics[width=5.0in]{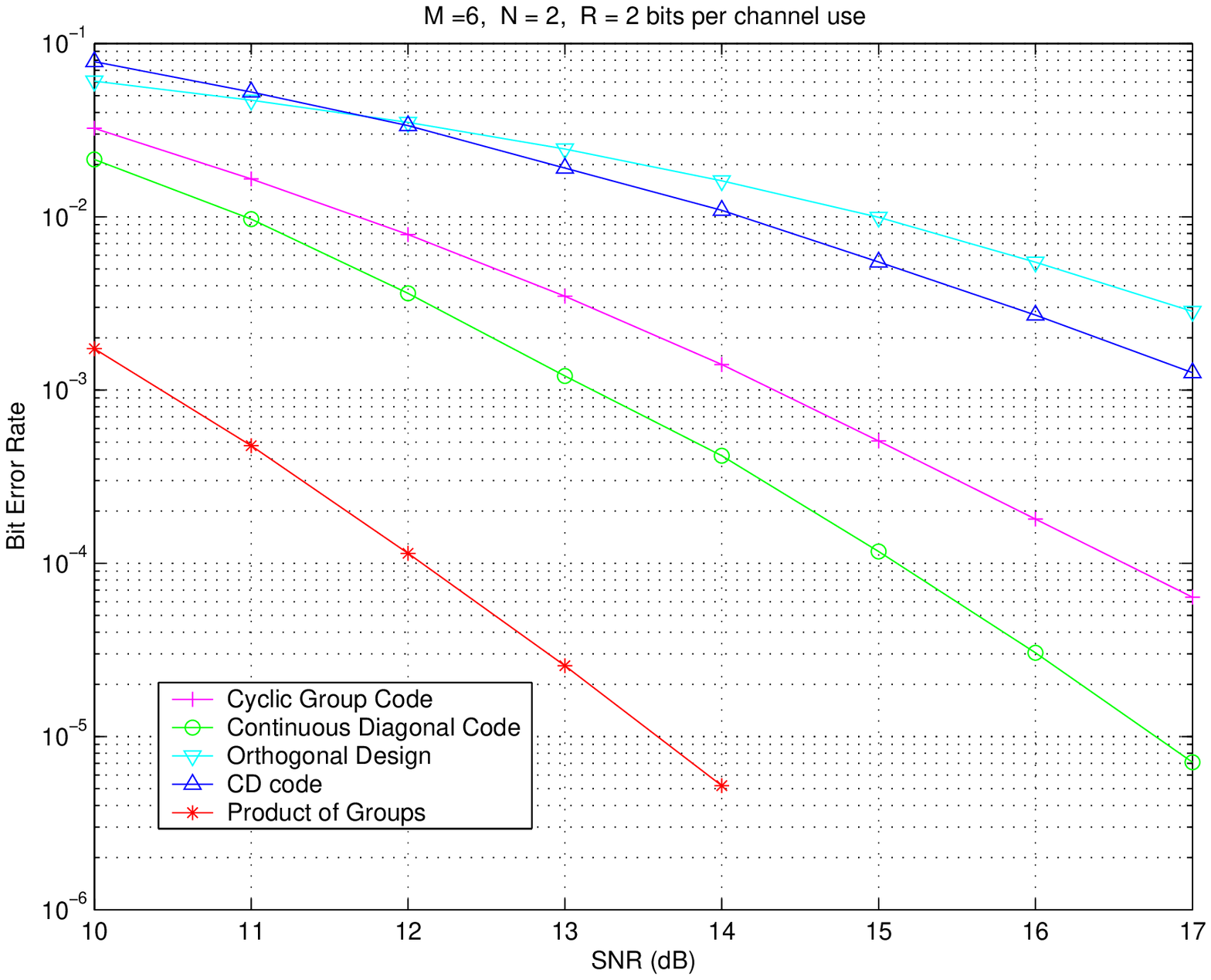}
\caption{Performance simulations of constellations} \label{fig58}
\end{figure*}

Specially, we have presented continuous diagonal codes for many combinations of antenna numbers and
constellation sizes in Table $1$ of Appendix \ref{app5}. These continuous diagonal codes outperform cyclic group
codes in terms of bit error rate.  For example, in Figure \ref{fig58}, with spectral efficiency $R =2$ bits per
channel use, our continuous diagonal code $\Lambda = [1 \;\; 11.8659 \;\; 404.3640 \;\; 592.2112 \;\; 1328.7582
\;\; 1489.9040]$ improves upon the best previously known cyclic group code $u = [1 \; 599 \; 623 \; 1445 \; 1527
\; 1715]$ (see Table I of \cite{Shokrollahi}) about $1.5$ dB at bit error rate $10^{-4}$ when using $6$
transmitter antennas and $2$ receiver antennas. In Figure \ref{fig58}, it shown that our continuous diagonal
code also substantially outperforms the orthogonal design and the CD code (with $(Q, \xi) = (12, 0.5401)$ as
mentioned before). However, the product of groups code has much better performance than our continuous diagonal
code. For the product of groups code, we used the same diagonal elements, $u = [1, \; 9, \; 21, \; 51, \; 53, \;
57]$, as that of \cite{Shokrollahi}. We searched the best $T$ matrix based on the conventional criterion of
diversity product maximization. We obtained a $T$ matrix so that the constellation has diversity product
$0.2098$, which is greater than the previously known value, $0.2084$, reported in \cite{Shokrollahi}.

It is important to note that,  since the constellation size of many types of FPF codes is not a power of $2$,
the bit assignment is not trivial and may significantly increase bit error rate.  The first method of bit
assignment is to truncate the constellation as a smaller one so that the size is of a power of $2$. The drawback
with this mapping method is that a large portion of the signal matrices may be wasted.  For example, suppose we
have an optimal (or near optimal) constellation  of $240$ signal matrices but only $128$ of them is used to
convey information. One can argue that it may be better to directly seek the optimal (or near optimal)
constellation of $128$ signal matrices. The second method is to map $n$ consecutive bits into $m$ consecutively
transmitted matrices where $m$ and $n$ are integers large enough so that $2^n$ is close to the $m$-th power of
constellation size (see, pp. 2356-2357 of \cite{Shokrollahi}). Unfortunately, the bit error rate will be
increased as the product of the block error rate and $ \eta m$ where $\eta \in (0,1)$ may not be small.
Moreover, the decoding delay is increased as $m M$ symbol periods, which may be intolerable for large $m$ and
$M$. The increase of bit error rate and decoding delay can be substantial since the factor $m$ can be quite
large. For example, when the constellation size is $240$, the minimal values of integer $m$ to guarantee $1 <
\frac{240^m}{2^n} \leq 1.05$ and $1 < \frac{240^m}{2^n} \leq 1.01$ are respectively $10$ and $118$. It can be
seen from the above analysis that for practical purpose the size of signal constellation should be a power of
$2$. This is one of the reasons why we choose code parameters to be continuous so that we can find unitary
space-time constellations of any size.

Finally, we would like to point out that some of the constellations we obtained have {\it zero} diversity
product. However, these constellations significantly outperform other constellations with much larger diversity
product. Such constellations can be found in Appendix \ref{app5} for the following combinations: (i) $M=N=2, \;
b=2$; (ii) $M=N=2, \; b=3$; (iii) $M=4, \; N=2, \; b =2$. As comparing to diversity product, our computational
experiments indicate that the trapezoid criterion introduced in Section 2 works quite well even in low SNR
region.

\bigskip

\bigskip

\bigskip

\bigskip

\section{Conclusion} \label{conclusion}

We have proposed a new class of differential unitary space-time
codes which has high performance, low encoding and decoding
complexity. We have established a parallel sphere decoder algorithm
which efficiently decodes our proposed code and existing codes such
as cyclic group code, FPF code $G_{m,r}$, non-group code $S_{m,s}$
and products of groups. We have proposed a new design criterion and
powerful optimization techniques for designing unitary space-time
codes. We have obtained constellations which significantly improve
upon constellations reported in the literature.

\appendix

\section{PROOF OF THEOREM \ref{th_diag}} \label{app1}

From the illustration after Theorem \ref{them2}, we see that, the
Chernoff bound of the pair-wise error probability is invariant
under unitary transforms.  By such invariant property, we have
\[
P( \Lambda_p^{\ell}, \Lambda_p^{\ell^{'}  } ) = P( I,
\Lambda^{\ell^{'} - \ell  } ),  \;\;\;\;\; 0 \leq p \leq 2^b-1.
\]
Note that
\[
d^\mathrm{H}(p, p, \ell, \ell^{'}) = d^\mathrm{H}(\ell, \ell^{'})
\]
for $0 \leq p \leq 2^b-1$.   Hence
\begin{eqnarray}
&    & \sum_{p =0}^{2^b - 1} \; \sum_{\ell =0}^{L - 2} \;
\sum_{\ell^{'} = \ell+1}^{L - 1} d^\mathrm{H}(p, p, \ell,
\ell^{'}) \; P( \Lambda_p^{\ell},
\Lambda_p^{\ell^{'}  } ) \nonumber\\
& = & 2^b \; \sum_{\ell =0}^{L - 2} \; \sum_{\ell^{'} = \ell+1}^{L
- 1} d^\mathrm{H}( \ell, \ell^{'}) \; P( I, \Lambda^{\ell^{'} -
\ell } )
\nonumber\\
& = & 2^b \;  \sum_{k=1}^{L-1} \; \sum_{ \ell^{'} - \ell = k
\atop{ 0 \leq \ell \leq L-2 \atop{\ell^{'} \leq L-1 }  }  }
d^\mathrm{H}( \ell^{'} , \ell) \; P( I, \Lambda^k ). \label{th2_0}
\end{eqnarray}
It can be verified that
\begin{eqnarray}
\sum_{ \ell^{'} - \ell = k \atop{ 0 \leq \ell \leq L-2
\atop{\ell^{'} \leq L-1 }  }  } d^\mathrm{H}( \ell^{'} , \ell) & =
& \sum_{0\leq \ell \leq L-k-1} \;
d^\mathrm{H}( \ell+k , \ell) \nonumber\\
& = & w(k).  \label{th2_1}
\end{eqnarray}
By (\ref{th2_0}) and (\ref{th2_1}), \begin{eqnarray} &   &  \sum_{p =0}^{2^b - 1} \; \sum_{\ell =0}^{L - 2} \;
\sum_{\ell^{'}  = \ell+1}^{L - 1} d^\mathrm{H}(p, p, \ell, \ell^{'}) \; P( \Lambda_p^{\ell}, \Lambda_p^{\ell^{'}
} )\nonumber\\
& = & 2^b \; \sum_{k=1}^{L-1} \; w(k) \; P( I, \Lambda^k ). \label{eqb} \end{eqnarray} Observing that
\[
d^\mathrm{H}(p, q, \ell, \ell^{'}) = d^\mathrm{H}(\ell, \ell^{'})
+ d^\mathrm{H}(p, q)
\]
and
\[
P( A_p \Lambda_p^{\ell} B_p, \; A_q \Lambda_q^{\ell^{'}} B_q) = P(
B_p, \; \Lambda^{\ell^{'} - \ell} B_q),
\]
we have
\begin{eqnarray}
&    & \sum_{ \ell = 0}^{  L - 1  } \; \sum_{ \ell^{'} = 0}^{ L -
1  }  d^\mathrm{H}(p, q, \ell, \ell^{'}) \;
P( A_p \Lambda_p^{\ell} B_p, \; A_q \Lambda_q^{\ell^{'}} B_q) \nonumber\\
& = &  \sum_{ \ell = 0}^{  L - 1  } \; \sum_{ \ell^{'} = 0}^{ L - 1  } d^\mathrm{H}(\ell, \ell^{'}) \; P( B_p,
\; \Lambda^{\ell^{'} - \ell} B_q) \nonumber\\
&  &   + \; \sum_{ \ell = 0}^{  L - 1
  } \;
\sum_{ \ell^{'} = 0}^{ L - 1  } d^\mathrm{H}(p, q) \;
P( B_p, \; \Lambda^{\ell^{'} - \ell} B_q) \nonumber \\
& = & \sum_{k=-L+1}^{L-1} \sum_{ \ell^{'} - \ell = k \atop{0 \leq \ell \leq L-1 \atop{ 0 \leq \ell^{'} \leq L-1
} } } d^\mathrm{H}(\ell, \ell^{'})  \; P( B_p, \; \Lambda^k B_q) \nonumber\\
&  &  +
  \; d^\mathrm{H}(p, q) \; \sum_{k=-L+1}^{L-1} P( B_p, \; \Lambda^k B_q)
\label{eq8888}.
\end{eqnarray}
Making use of symmetry, we can show that \be \sum_{ \ell^{'} - \ell = k \atop{0 \leq \ell \leq L-1 \atop{ 0 \leq
\ell^{'} \leq L-1   } } } d^\mathrm{H}(\ell, \ell^{'}) = w(|k|). \label{th2_2} \ee By (\ref{eq8888}) and
(\ref{th2_2}),
\begin{eqnarray}
&    & \sum_{ \ell = 0}^{  L - 1  } \; \sum_{ \ell^{'} = 0}^{ L -
1  }  d^\mathrm{H}(p, q, \ell, \ell^{'}) \;
P( A_p \Lambda_p^{\ell} B_p, \; A_q \Lambda_q^{\ell^{'}} B_q) \nonumber\\
&  = & \sum_{k=-L+1}^{L-1} \; [w(|k|) + d^\mathrm{H}(p, q)] \; P(
B_p, \; \Lambda^k B_q) \label{eq38}.
\end{eqnarray}
The proof is finally completed by invoking
equations~(\ref{eqbit}), ~(\ref{eqb}) and ~(\ref{eq38}).

\bigskip

\section{PROOF OF THEOREM \ref{them2}} \label{app2}

\bigskip

By virtue of the Chernoff bound (\ref{chernoff}),
\begin{eqnarray*}
&  & P (U, \Phi) \\
& = & \frac{1}{2} \prod_{m=1}^M \left[ 1 +
\frac{\rho^2 \sigma_m^2 }{4( 1 + 2 \rho) } \right]^{-N}\\
& = & \frac{\alpha ^{MN}}{2}  \left[\prod_{m=1}^M \left( \alpha +
\sigma_m^2  \right)\right]^{-N}
\end{eqnarray*}
where $\sigma_m$ is the $m$-th singular value of $U - \Phi$. Let
$U_1$ and $U_2$ be unitary matrices such that $ U - \Phi = U_1  \;
{\rm diag} (\sigma_1, \cdots, \sigma_M) \; U_2^{\dag}$. Then
\begin{eqnarray}
&    & \det[\alpha I + (U - \Phi) (U - \Phi)^\dag  ] \nonumber\\
 & = & \det [ \alpha I +  U_1 \; {\rm diag} (\sigma_1^2, \cdots,
\sigma_M^2)  \;
U_1^{\dag}] \nonumber\\
& = & \det [ U_1 \; {\rm diag} (\alpha + \sigma_1^2, \cdots,
\alpha +
\sigma_M^2)  \; U_1^{\dag}] \nonumber\\
& = & \det( U_1 U_1^\dag) \; \det [ {\rm diag} (\alpha +
\sigma_1^2,
\cdots, \alpha + \sigma_M^2)] \nonumber\\
& = & \prod_{m = 1}^M (\alpha + \sigma_m^2).  \label{th3_0}
\end{eqnarray}
It follows that \be P (U, \Phi)  = \frac{\alpha^{MN}} {2 \left(
\det[ \alpha I + (U - \Phi) (U - \Phi)^{\dag}  ]  \right)^N }
\label{identity} \ee from which we obtain
\begin{eqnarray*}
P (I, \Phi) & = & \frac{\alpha^{MN}} {2 \left( \det[ \alpha I
+ (I - \Phi) (I - \Phi)^{\dag}  ]  \right)^N }\\
& = & \frac{ \alpha^{MN}} {2 \left( \det[ (\alpha + 2) I - \Phi -
\Phi^\dag  ]  \right)^N }
\end{eqnarray*}
by letting $U = I$.  This proves (\ref{union_0}).

\bigskip
Now define
\begin{eqnarray}
\Xi & \stackrel{\mathrm{def}}{=} & \log \det[\alpha I + (U - \Phi)
(U -
\Phi)^{\dag}] \nonumber\\
& = & \log \det [(\alpha + 2) I - U \Phi^\dag - \Phi U^\dag].
\label{th3_1}
\end{eqnarray}
By the chain rule of differentiation,
\begin{eqnarray}
\frac{ \partial P (U , \Phi) } { \partial \nu_{pq}   } & = &
\frac{ \partial P (U , \Phi) } { \partial \Xi   } \;
\frac{ \partial \Xi } { \partial \nu_{pq}   } \nonumber\\
& = & \frac{ \partial \left[ \frac{\alpha^{MN} }{2}\exp(- N \Xi)
\right]} {
\partial \Xi   } \;
\frac{ \partial \Xi } { \partial \nu_{pq}   } \nonumber \\
& = & - N  P(U , \Phi) \; \frac{ \partial \Xi} { \partial \nu_{pq}
}. \label{th3_2}
\end{eqnarray}
Similarly, \be \frac{ \partial P (U , \Phi) } { \partial \phi_{pq}
} = - N  P(U , \Phi) \; \frac{ \partial \Xi} { \partial \phi_{pq}
}, \label{th3_333} \ee \be \frac{ \partial P (U , \Phi) } {
\partial \theta_{k} } = - N  P (U , \Phi) \; \frac{
\partial \Xi} {
\partial \theta_{k} }. \label{th3_3} \ee Define
\[
\Omega \stackrel{\mathrm{def}}{=} (\alpha + 2) I - U \Phi^\dag -
\Phi U^\dag.
\]
By (\ref{th3_0}) and (\ref{th3_1}),
\[
\det(\Omega) = \prod_{m = 1}^M (\alpha + \sigma_m^2) \geq \alpha
^M  > 0
\]
for any $U$.  Let $e_j$ be the $M$-dimensional unit column vector
with a one in the $j$-th entry and zeros elsewhere.  By the same
method of proving (\ref{th3_0}), we can show that $\det [(\alpha +
2) I - (U + e_j e_k^{\intercal} \delta ) \Phi^\dag - \Phi (U^\dag
+ e_k e_j^{\intercal} \delta )]$ is a positive real number for any
$\delta \in \mathbb{R}$. Since $\det(\Omega)$ is positive and
\begin{eqnarray*}
&   & \det [(\alpha + 2) I - (U + e_j e_k^{\intercal} \delta ) \Phi^\dag - \Phi (U^\dag + e_k e_j^{\intercal}
\delta
)]\\
& = & \det(\Omega) \det[I - \Omega^{-1} (e_j e_k^{\intercal} \Phi^\dag + \Phi e_k e_j^{\intercal}) \delta  ],
\end{eqnarray*}
we have that $\det[I - \Omega^{-1} (e_j e_k^{\intercal} \Phi^\dag + \Phi e_k e_j^{\intercal}) \delta  ]$ is also
a positive real number for any $\delta \in \mathbb{R}$.  Therefore,
\begin{eqnarray*}
&   & \log \det [(\alpha + 2) I -
(U + e_j e_k^{\intercal} \delta ) \Phi^\dag - \Phi (U^\dag + e_k e_j^{\intercal} \delta )]\\
& = & \log \det(\Omega) + \log \det[I - \Omega^{-1} (e_j
e_k^{\intercal} \Phi^\dag + \Phi e_k e_j^{\intercal}) \delta  ].
\end{eqnarray*}
Let $\Psi = \Omega^{-1} (e_j e_k^{\intercal} \Phi^\dag + \Phi e_k
e_j^{\intercal})$.  Then $\Psi$ is a Hermite matrix, i.e.,
$\Psi^\dag = \Psi$. It follows that $[\Psi]_{kk}$ is real for $k =
1, \cdots, M$.   By the definition of a determinant, we have
\[
\det(I - \Psi \delta) = \prod_{k=1}^M (1 - [\Psi]_{kk} \delta) +
\delta^2 f(\delta) > 0
\]
where $f(.)$ is a polynomial function of $\delta \in \mathbb{R}$.
Since $\det(I - \Psi \delta)$ and $[\Psi]_{kk}, \; k = 1, \cdots,
M$ are real numbers, it must be true that $f(\delta)$ is also a
real-valued function of $\delta \in \mathbb{R}$.  Note that
\begin{eqnarray*}
\det(I - \Psi \delta) & = & 1 - \left( \sum_{k=1}^M [\Psi]_{kk}
\right
) \delta + O(\delta^2) + \delta^2 f(\delta)\\
& = &  1 - {\rm tr} (\Psi) \delta + O(\delta^2)\\
& > & 0
\end{eqnarray*}
where ${\rm tr} (\Psi)$ is real and $O(\delta^2)$ is a real-valued
function of $\delta \in \mathbb{R}$.  Therefore, \be \label{gentr}
 \log \det(I - \Psi \delta) = - {\rm tr} (\Psi)
\delta + O(\delta^2). \ee Making use of (\ref{gentr}), we have
\begin{eqnarray*}
&   & \log \det [(\alpha + 2) I -
(U + e_j e_k^{\intercal} \delta ) \Phi^\dag - \Phi (U^\dag + e_k e_j^{\intercal} \delta )]\\
& = & \log \det(\Omega) - {\rm tr} ( \Psi \delta) + O(\delta^2)\\
& = & \log \det(\Omega)\\
&  &  - \left[{\rm tr} \left( e_j e_k^{\intercal} \Phi^\dag \Omega^{-1} \right)  + {\rm tr} \left( \Omega^{-1}
\Phi e_k e_j^{\intercal} \right) \right] \delta   +
O(\delta^2)\\
& = & \log \det(\Omega)\\
&   &  - \left[{\rm tr} \left( e_j e_k^{\intercal} ( \Omega^{-1} \Phi )^\dag \right)  + {\rm tr} \left(
\Omega^{-1} \Phi e_k e_j^{\intercal} \right) \right] \delta +
O(\delta^2)\\
& = & \log \det(\Omega)\\
&  &  - \left[ {\rm tr} (\left( \Omega^{-1} \Phi e_k e_j^{\intercal} \right)^\dag) + {\rm tr} \left( \Omega^{-1}
\Phi e_k e_j^{\intercal} \right) \right] \delta   +
O(\delta^2)\\
& = & \log \det(\Omega)\\
&  &  - \left[ ( {\rm tr} \left( \Omega^{-1} \Phi e_k e_j^{\intercal} \right) )^\dag + {\rm tr} \left(
\Omega^{-1} \Phi e_k e_j^{\intercal} \right) \right] \delta   +
O(\delta^2)\\
& = & \log \det(\Omega) - 2 \; \Re ( {\rm tr} \left( \Omega^{-1}
\Phi
e_k e_j^{\intercal} \right)  ) \; \delta   + O(\delta^2)\\
& = & \log \det(\Omega) - 2 \; \Re ( [\Omega^{-1} \Phi ]_{jk} ) \;
\delta
+ O(\delta^2)\\
& = & \log \det(\Omega) - 2 \; [\Re (\Omega^{-1} \Phi)]_{jk}  \;
\delta   + O(\delta^2)
\end{eqnarray*}
for any $\delta \in \mathbb{R}$. Therefore,  applying formula
\[
\left[ \frac{ \partial \; f(X)} { \partial \; \Re(X)} \right]_{jk}
= \lim_{\delta \rightarrow 0} \frac{ f(X + e_j e_k^\intercal
\delta) - f(X)  } {\delta}
\]
provided in \cite{Hassibi1} (page 1501), we have
\begin{eqnarray*}
&  & \left[ \frac{ \partial \; \Xi} { \partial \; \Re(U)} \right]_{jk}\\
 & = & \lim_{\delta \rightarrow 0} \frac{
\log \det [\Omega - (e_j e_k^{\intercal} \Phi^\dag + \Phi e_k e_j^{\intercal}) \delta ] -
\log \det(\Omega)  } {\delta}\\
& = & \lim_{\delta \rightarrow 0} \frac{ - 2 [\Re (\Omega^{-1}
\Phi)]_{jk}  \; \delta   + O(\delta^2)  }
{\delta}\\
& = &  - 2 [\Re (\Omega^{-1} \Phi)]_{jk}.
\end{eqnarray*}
Observing that $U = I$ for $\Theta = 0$ (i.e., all elements of $\Theta$ are zeros), we have $\Omega = (\alpha +
2) I - \Phi - \Phi^\dag$ and $\Omega^{-1} \Phi  =  [(\alpha + 2) I - \Phi - \Phi^{\dag}]^{-1} \Phi =
\mathcal{Q}$.
  Hence,
\be \left. \frac{ \partial \; \Xi } { \partial \; \Re(U) } \right|_{\Theta = 0} = - 2 \Re (\mathcal{Q}).
\label{th3_4} \ee Similarly,
\begin{eqnarray*}
&   & \log \det [(\alpha + 2) I -
(U + e_j e_k^{\intercal} \delta i ) \Phi^\dag - \Phi (U^\dag - e_k e_j^{\intercal} \delta i )]\\
& = & \log \det [\Omega - (e_j e_k^{\intercal} \Phi^\dag - \Phi e_k e_j^{\intercal}) \delta i ]\\
& = & \log \det(\Omega) + \log \det[I - \Omega^{-1} (e_j
e_k^{\intercal} \Phi^\dag - \Phi
e_k e_j^{\intercal}) \delta i ]\\
& = & \log \det(\Omega) - {\rm tr} [\Omega^{-1} (e_j
e_k^{\intercal} \Phi^\dag - \Phi
e_k e_j^{\intercal}) \delta i ] + O(\delta^2)\\
& = & \log \det(\Omega)\\
&  &  - \left[{\rm tr} \left( e_j e_k^{\intercal} \Phi^\dag \Omega^{-1} \right)  - {\rm tr} \left( \Omega^{-1}
\Phi e_k e_j^{\intercal} \right) \right] \delta i  +
O(\delta^2)\\
& = & \log \det(\Omega)\\
&  &  - \left[{\rm tr} \left( e_j e_k^{\intercal} ( \Omega^{-1} \Phi )^\dag \right)  - {\rm tr} \left(
\Omega^{-1} \Phi e_k e_j^{\intercal} \right) \right] \delta i  +
O(\delta^2)\\
& = & \log \det(\Omega)\\
&   &  - \left[ {\rm tr} (\left( \Omega^{-1} \Phi e_k e_j^{\intercal} \right)^\dag) - {\rm tr} \left(
\Omega^{-1} \Phi e_k e_j^{\intercal} \right) \right] \delta i  +
O(\delta^2)\\
& = & \log \det(\Omega)\\
&   &  - \left[ ( {\rm tr} \left( \Omega^{-1} \Phi e_k e_j^{\intercal} \right) )^\dag - {\rm tr} \left(
\Omega^{-1} \Phi e_k e_j^{\intercal} \right) \right] \delta  i +
O(\delta^2)\\
& = & \log \det(\Omega) - 2 \; \Im ( {\rm tr} \left( \Omega^{-1}
\Phi
e_k e_j^{\intercal} \right)  ) \; \delta   + O(\delta^2)\\
& = & \log \det(\Omega) - 2 \; \Im ( [\Omega^{-1} \Phi ]_{jk} ) \;
\delta
+ O(\delta^2)\\
& = & \log \det(\Omega) - 2 \; [\Im (\Omega^{-1} \Phi)]_{jk}  \;
\delta   + O(\delta^2)
\end{eqnarray*}
for any $\delta \in \mathbb{R}$.  Therefore,
\begin{eqnarray*}
&   & \left[ \frac{ \partial \; \Xi} { \partial \; \Im(U)} \right]_{jk}\\
 & = & \lim_{\delta \rightarrow 0} \frac{
\log \det [\Omega - (e_j e_k^{\intercal} \Phi^\dag - \Phi e_k e_j^{\intercal}) \delta i ] -
\log \det(\Omega)  } {\delta}\\
& = & \lim_{\delta \rightarrow 0} \frac{ - 2 [\Im (\Omega^{-1}
\Phi)]_{jk}  \; \delta   + O(\delta^2)  }
{\delta}\\
& = &  - 2 [\Im (\mathcal{Q})]_{jk}
\end{eqnarray*}
for $\Theta = 0$,  which implies that \be \left. \frac{ \partial \; \Xi } { \partial \; \Im(U) } \right|_{\Theta
= 0} = - 2 \Im (\mathcal{Q}). \label{th3_5} \ee

\bigskip

We now consider the partial derivatives of $U$ with respective to
the elements of $\Theta$.  It should be noted that an incorrect
formula for computing ${\left. \frac{ \partial U } { \partial
\phi_{pq}  } \right|} _{\Theta = 0}$ has been reported in
\cite{Agrawal} (see equation (13), page 2625). In the sequel, we
shall prove that \be {\left. \frac{ \partial U } { \partial
\phi_{pq}  } \right|} _{\Theta = 0} = e_q e_p^{\intercal} - e_p
e_q^{\intercal}, \label{th3_6} \ee which is clearly different from
equation (13) of \cite{Agrawal}.  To that end, we can use the
parameterization of unitary matrix $U(\Theta)$ to verify that
\[
{\left. \frac{ \partial U } { \partial \phi_{pq}  } \right|}
_{\Theta = 0} = {\left. \frac{ \partial \widetilde{U} } { \partial
\phi_{pq}  } \right|} _{\phi_{pq} = 0}
\]
where
\[
\widetilde{U} =  \left[\begin{array}{ll}
I_{ (p-1) \times (p-1) } & 0_{ (p-1) \times (M - p + 1) }\\
0_{ (M - p + 1) \times (p-1) } &  U^{p,q}
(\phi_{pq},0)\end{array}\right]
\]
with
\begin{eqnarray*}
&   & [U^{p,q} (\phi_{pq},0) ]_{jk}\\
&  =  & \left\{\begin{array}{ll}
1, \;\;\;&  {\rm if}\; j = k \; {\rm and} \;  j \notin \{1, \; q-p+1\}\\
\cos( \phi_{pq}  ), \;\;\;&  {\rm if}\; j = k \; {\rm and} \;  j
\in \{1, \;
q-p+1\}\\
- \sin( \phi_{pq}  ), \;\;\;&  {\rm if}\; j = 1 \; {\rm and} \;  k = q-p+1\\
\sin( \phi_{pq}  ), \;\;\;&  {\rm if}\; k = 1 \; {\rm and} \;  j = q-p+1\\
0, \; \;\;\;&  {\rm otherwise.}
\end{array} \right.
\end{eqnarray*}
Obviously,
\begin{eqnarray*}
&   & \left . \frac{ \partial \; (  [U^{p,q} (\phi_{pq},0) ]_{jk}) } {
\partial \phi_{pq} } \right |_{\phi_{pq} = 0}\\
&  = & \left\{\begin{array}{ll}
- 1, \;\;\;&  {\rm if}\; j = 1 \; {\rm and} \;  k = q-p+1\\
1, \;\;\;&  {\rm if}\; k = 1 \; {\rm and} \;  j = q-p+1\\
0, \; \;\;\;&  {\rm otherwise.}
\end{array} \right.
\end{eqnarray*}
Hence, (\ref{th3_6}) can be obtained by observing that
\[
[\widetilde{U}]_{pq} = [U^{p,q} (\phi_{pq},0) ]_{1, q-p+1}, \]
\[
 [\widetilde{U}]_{qp} = [U^{p,q} (\phi_{pq},0)
]_{q-p+1,1}.
\]
To compute other partial derivatives of $U(\Theta)$ at $\Theta =
0$, we quote equations (14) and (15) of \cite{Agrawal} as follows:
\be {\left. \frac{ \partial U } { \partial \nu_{pq}  } \right|}
_{\Theta = 0} = 0,   \;\;\;\;\; {\left. \frac{ \partial U } {
\partial \theta_{k}  } \right|} _{\Theta = 0} = ie_k
e_k^{\intercal}. \label{th3_7} \ee By virtue of (\ref{th3_6}) and
(\ref{th3_7}), \be {\left. \frac{ \partial \; \Re(U) } { \partial
\phi_{pq}  } \right|} _{\Theta = 0} = e_q e_p^{\intercal} - e_p
e_q^{\intercal}, \;\;\;\;\;\;  {\left. \frac{ \partial \; \Im(U) }
{ \partial \phi_{pq}  } \right|} _{\Theta = 0} = 0, \label{th3_8}
\ee \be {\left. \frac{ \partial \; \Re(U) } {
\partial \nu_{pq}  } \right|} _{\Theta = 0} = 0,
\;\;\;\;\;\;\;\; \;\;\;\;\;\;\;\;\;\;\;\;\;  {\left. \frac{
\partial \; \Im(U) } { \partial \nu_{pq}  } \right|} _{\Theta =
0} = 0, \label{th3_9} \ee \be {\left. \frac{ \partial \; \Re(U) }
{ \partial \theta_{k}  } \right|} _{\Theta = 0} = 0,
\;\;\;\;\;\;\;\;\;\;\;\; \;\;\;\;\; {\left. \frac{
\partial \; \Im(U) } { \partial \theta_{k}  } \right|} _{\Theta =
0} = e_k e_k^{\intercal}. \label{th3_10} \ee

\bigskip

We now define inner product $<.,.>$ by
\[
<X,Y> \stackrel{\mathrm{def}}{=} \sum_{j,k} [X]_{j k} \; [Y]_{j
k}.
\]
Then by the chain rule of differentiation and equations (\ref{th3_4}), (\ref{th3_8}), we have
\begin{eqnarray*}
\frac{ \partial \; \Xi} { \partial \phi_{pq}  } & = & \left<
\frac{ \partial \; \Xi } { \partial \; \Re(U)  }, \frac{ \partial
\; \Re(U) } { \partial \phi_{pq}  } \right> + \left< \frac{
\partial \; \Xi } { \partial \; \Im(U)  }, \frac{ \partial \;
\Im(U) }
{ \partial \phi_{pq}  } \right>\\
& = & \left< \frac{ \partial \; \Xi } { \partial \; \Re(U)  },
\frac{ \partial \; \Re(U) }
{ \partial \phi_{pq}  } \right>\\
& = & \left< -2 \; \Re(\mathcal{Q}), \; e_q e_p^{\intercal} - e_p e_q^{\intercal}
\right>\\
& = & - 2  \; \Re \left( \left< \mathcal{Q}, \; e_q e_p^{\intercal} - e_p
e_q^{\intercal} \right> \right)\\
& = & - 2 \; \Re( [\mathcal{Q}]_{q p} - [\mathcal{Q}]_{p q} ).
\end{eqnarray*}
Invoking (\ref{th3_333}) yields
\begin{eqnarray*}
\frac{ \partial \; P(U, \Phi)} { \partial \phi_{pq}  } & = & - N
P(U, \Phi) \frac{ \partial \; \Xi} { \partial
\phi_{pq}  }\\
& = & 2 N P(U, \Phi) \; \Re( [\mathcal{Q}]_{q p} - [\mathcal{Q}]_{p q} )
\end{eqnarray*}
and hence proves (\ref{grad_0}).

By the chain rule of differentiation and (\ref{th3_9}), \be \frac{
\partial \; \Xi } { \partial \nu_{pq}  } = \left< \frac{
\partial \; \Xi } { \partial \; \Re(U)  }, \frac{ \partial \;
\Re(U) } { \partial \nu_{pq}  } \right> + \left< \frac{
\partial \; \Xi } { \partial \; \Im(U)  }, \frac{ \partial \;
\Im(U) } { \partial \nu_{pq}  } \right> = 0. \label{th3_11} \ee
Combing (\ref{th3_2}) and (\ref{th3_11}) leads to
\[
\frac{ \partial \; P(U, \Phi)} { \partial \nu_{pq}  }  = 0
\]
and thus completes the proof of (\ref{grad_2}).

By the chain rule of differentiation and (\ref{th3_10}),
\begin{eqnarray*}
\frac{ \partial \; \Xi } { \partial \theta_{k}  } & = & \left<
\frac{ \partial \; \Xi } { \partial \; \Re(U)  }, \frac{ \partial
\; \Re(U) } { \partial \theta_{k}  } \right> + \left< \frac{
\partial \; \Xi } { \partial \; \Im(U)  }, \frac{ \partial \;
\Im(U) }
{ \partial \theta_{k}  } \right>\\
& = & \left< \frac{ \partial \; \Xi } { \partial \; \Im(U)  },
\frac{ \partial \; \Im(U) }
{ \partial \theta_{k}  } \right>\\
& = & \left < - 2 \; \Im (\mathcal{Q}) , \;\;  e_k e_k^{\intercal} \right > \\
& = & - 2 \; \Im \left( \left < \mathcal{Q}, \;\;  e_k e_k^{\intercal} \right >
\right)\\
& = & - 2 \; \Im( [\mathcal{Q}]_{k k} ).
\end{eqnarray*}
Hence by (\ref{th3_3}), we have
\begin{eqnarray*}
\frac{ \partial \; P(U, \Phi)} { \partial \theta_{k}  } & = & - N
P(U, \Phi) \; \frac{ \partial \; \Xi} { \partial
\theta_{k}  }\\
& = & 2 N P(U, \Phi) \; \Im( [\mathcal{Q}]_{k k} )
\end{eqnarray*}
and completes the proof of (\ref{grad_1}).

\bigskip

Define
\[
\Upsilon \stackrel{\mathrm{def}}{=} \log \det[\alpha I +
(\Lambda^\ell - \Phi) (\Lambda^\ell - \Phi)^{\dag}].
\]
By the same method as computing $\frac{ \partial \; \Xi } {
\partial \; \Im(U)  }$, we have \be \frac{ \partial \; \Upsilon }
{ \partial \; \Im(\Lambda^\ell)  } = - 2 \Im (\mathcal{Q}). \label{th3_12} \ee Observing that
$[\Lambda^\ell]_{jk}$ depends on $\lambda_m$ only if $j = k = m$ and that
\[
{\left. \frac{ \partial \; \Re([\Lambda^\ell]_{mm}) } { \partial
\lambda_m  } \right|} _{\Lambda = I} = {\left. \frac{ \partial \;
\cos(\frac{2 \pi \ell \lambda_m } {L} ) } { \partial \lambda_m  }
\right|} _{\Lambda = I} =  0
\]
\[
{\left. \frac{ \partial \; \Im([\Lambda^\ell]_{mm}) } { \partial
\lambda_m  } \right|} _{\Lambda = I} = {\left. \frac{ \partial \;
\sin(\frac{2 \pi \ell \lambda_m } {L} ) } { \partial \lambda_m  }
\right|} _{\Lambda = I} =  \frac{2 \pi \ell}{L},
\]
we have \be {\left. \frac{ \partial \; \Re(\Lambda^\ell) } {
\partial \lambda_m  } \right|} _{\Lambda = I} = 0, \;\;\; {\left.
\frac{
\partial \; \Im(\Lambda^\ell) }
{ \partial \lambda_m  } \right|} _{\Lambda = I} = \frac{2 \pi
\ell}{L} e_m e_m^{\intercal}. \label{th3_13} \ee By the chain rule
of differentiation and equations (\ref{th3_12}), (\ref{th3_13}),
we have
\begin{eqnarray*}
\frac{ \partial \; \Upsilon } { \partial \lambda_m  } & = & \left<
\frac{ \partial \; \Upsilon } { \partial \; \Re(\Lambda^\ell)  },
\frac{ \partial \; \Re(\Lambda^\ell) } { \partial \lambda_m  }
\right> + \left< \frac{ \partial \; \Upsilon } { \partial \;
\Im(\Lambda^\ell)  }, \frac{ \partial \; \Im(\Lambda^\ell) } {
\partial \lambda_m  } \right>\\
& = & \left< \frac{ \partial \; \Upsilon } { \partial \;
\Im(\Lambda^\ell)  }, \frac{ \partial \; \Im(\Lambda^\ell) } {
\partial \lambda_m  } \right>\\
& = & \left<  - 2 \Im(\mathcal{Q}), \frac{2 \pi \ell}{L} e_m e_m^{\intercal} \right>\\
& = & - \frac{4 \pi \ell}{L} \Im ( [\mathcal{Q}]_{mm} ).
\end{eqnarray*}
It follows that
\begin{eqnarray*}
\frac{ \partial \; P(\Lambda^\ell, \Phi)} { \partial \lambda_{m} }
& = & - N P(\Lambda^\ell, \Phi) \;  \frac{
\partial \; \Upsilon} {
\partial \lambda_m  }\\
& = & \frac{4 \pi \ell N P(\Lambda^\ell, \Phi) }{L} \; \Im( [\mathcal{Q}]_{m m} )
\end{eqnarray*}
and (\ref{grad_3}) is true.

\bigskip

\section{PROOF OF THEOREM \ref{reduce}} \label{app3}

\bigskip

First we need to prove some preliminary results.

\bigskip

\begin{lemma}  \label{left}
For any $\ell \in \{ 0, 1, \cdots,  L-1 \}$, there exists $y \in \mathcal{ S}$ such that \be \sum_{m =1}^M [(C_m
\lambda_m \; \ell - C_m \varphi_m) \; {\rm mod}^* C_m L]^2 = || y \; G - \xi ||^2 \label{equlem11}. \ee
\end{lemma}

\bigskip

\begin{pf}

Given $\ell \in \{0,1,\cdots, L-1\}$, define \be y_1 = \ell +
\left \lceil \frac{ \varphi_1 - \ell } {L} - \frac{1}{2} \right
\rceil L. \label{eqa_app_0} \ee We claim that \be - \frac{L}{2} +
\varphi_1 \leq y_1 < \frac{L}{2} + \varphi_1. \label{eqa_app_1}
\ee To prove (\ref{eqa_app_1}), one can make use of the
observation that
\[
0 \leq \lceil x \rceil - x  < 1 \;\; \forall x \in \mathbb{R}
\]
and verify  that  inequality
\[
0 \leq \left \lceil \frac{ \varphi_1 - \ell } {L} - \frac{1}{2}
\right \rceil - \left ( \frac{ \varphi_1 - \ell } {L} -
\frac{1}{2} \right ) < 1
\]
is equivalent to
\[
- \frac{L}{2} + \varphi_1 \leq \ell + \left \lceil \frac{
\varphi_1 - \ell } {L} - \frac{1}{2} \right \rceil L < \frac{L}{2}
+ \varphi_1.
\]
The truth of (\ref{eqa_app_1}) allows one to choose $y = [y_1, \;\cdots, y_M] \in \mathcal{ S}$ such that the
first entry of $y$ is $y_1$.  Let \be w = [w_1, \cdots, w_M] = y \; G - \xi. \label{eqa_app_2} \ee Obviously, to
show (\ref{equlem11}), it suffices to show
\[
(C_m \lambda_m \; \ell - C_m \varphi_m) \; {\rm mod}^* C_m L =
w_m, \;\;\; m = 1,\cdots, M
\]
where $\lambda_1 = 1$.  By the definitions of $G$ and $\xi$,  we
can rewrite (\ref{eqa_app_2}) as
\[
w_1 = C_1 y_1 - C_1 \varphi_1, \] \[
w_m = C_m \lambda_m \; y_1 + C_m L \; y_m - C_m \varphi_m \;\; {\rm for}
\;\; m = 2, \cdots, M.
\]
Hence, to show (\ref{equlem11}), it suffices to show \be (C_1  \; \ell - C_1 \varphi_1) \; {\rm mod}^* C_1 L =
C_1 y_1 - C_1 \varphi_1 \label{eqy1} \ee and, for $m = 2, \cdots, M$, \begin{eqnarray} &  & (C_m \lambda_m \;
\ell - C_m \varphi_m) \; {\rm mod}^* C_m L \nonumber\\
&  = & C_m \lambda_m \; y_1 + C_m L \; y_m - C_m \varphi_m. \label{eqyms}
\end{eqnarray} Note that, for any $\ell$, there exits an unique integer $z_1$ such that
\[
(C_1  \; \ell - C_1 \varphi_1) \; {\rm mod}^* C_1 L =  C_1  \;
\ell - C_1 \varphi_1 + z_1 C_1 L.
\]
Therefore, to show (\ref{eqy1}), it suffices to show
\[
C_1 y_1 - C_1 \varphi_1 = C_1  \; \ell - C_1 \varphi_1 + z_1 C_1
L,
\]
or equivalently, \be z_1 = \frac{y_1 - \ell}{L}. \label{eqy} \ee By the definition of the symmetric modulus
operator ${\rm mod}^*$, integer $z_1$ guarantees
\[
- \frac{C_1 L}{2} \leq C_1  \; \ell - C_1 \varphi_1 + z_1 C_1 L <
\frac{C_1 L}{2},
\]
or equivalently,
\[
- 1 < \left( \frac{ \varphi_1 - \ell } { L } - \frac{1}{2} \right)
- z_1 \leq 0,
\]
which implies
\[
\left\lceil \frac{ \varphi_1 - \ell } { L } - \frac{1}{2} - z_1
\right \rceil = 0.
\]
Since $z_1$ is an integer, we have
\begin{eqnarray*}
z_1 & = & \left\lceil \frac{ \varphi_1 - \ell } { L } -
\frac{1}{2} \right
\rceil\\
& = & \frac{y_1 - \ell}{L}
\end{eqnarray*}
where the second equality follows from (\ref{eqa_app_0}). So
equation (\ref{eqy1}) is proven by invoking (\ref{eqy}).

\bigskip

In light of the fact that, for any given $\ell \in \{0,1,
\cdots,L-1 \}$ and for any $m \in \{2, \cdots, M\}$, there exists
an unique integer $z_m$ such that
\[
(C_m \lambda_m \; \ell - C_m \varphi_m) \; {\rm mod}^* C_m L = C_m (\lambda_m \ell - \varphi_m + z_m  L),
\]
to show (\ref{eqyms}) it suffices to prove that
\[
C_m (\lambda_m \ell - \varphi_m + z_m L) = C_m \lambda_m \; y_1 + C_m L \; y_m - C_m \varphi_m \] for $m=
2,\cdots, M$,  or equivalently, \be y_m = z_m - \frac{y_1 - \ell}{L} \lambda_m, \;\;\;\;\;\; m= 2,\cdots, M.
\label{eqa_app_3} \ee By the definition of the symmetric modulus operator ${\rm mod}^*$, integer $z_m$
guarantees
\[
- \frac{C_m L}{2} \leq C_m \lambda_m \; \ell - C_m \varphi_m + z_m
C_m L < \frac{C_m L}{2}
\]
which can be rewritten as
\[
- \frac{C_m L}{2} \leq C_m [\lambda_m ( \ell + z_1 L)- \varphi_m + (z_m - z_1 \lambda_m ) L ]< \frac{C_m L}{2},
\]
i.e.,
\[
-1 < \left( \frac{\varphi_m}{L} - \left ( \frac{\ell + z_1 L}{L} - z_1 \right ) \; \lambda_m - \frac{1}{2}
\right)  - z_m \leq 0 \] for $m = 2, \cdots, M$. Therefore,
\[
\left \lceil \left( \frac{\varphi_m}{L} - \left ( \frac{\ell + z_1 L}{L} - z_1 \right ) \; \lambda_m -
\frac{1}{2} \right)  - z_m \right \rceil = 0 \] for $m = 2, \cdots, M$. Since $z_m$ is an integer, we have
\begin{eqnarray}
z_m  & = & \left \lceil   \frac{\varphi_m}{L} - \left ( \frac{\ell
+ z_1 L}{L} - z_1 \right ) \;
\lambda_m - \frac{1}{2}     \right \rceil \nonumber\\
& = & \left \lceil   \frac{\varphi_m}{L} - \left ( \frac{y_1}{L} -
z_1 \right ) \; \lambda_m - \frac{1}{2}     \right \rceil
\label{eqa_app_4}
\end{eqnarray}
for $m = 2 , \cdots, M$. Here (\ref{eqa_app_4}) is due to (\ref{eqy}).  By the definition of $\mathcal{ S}$, \be
y_m = \left \lceil   \frac{\varphi_m}{L} - \left ( \frac{y_1}{L} - \left \lfloor \frac{y_1}{L} \right \rfloor
\right ) \; \lambda_m - \frac{1}{2}  \right \rceil - \left \lfloor \frac{y_1}{L} \right \rfloor \; \lambda_m.
\label{eqa_app_5} \ee By virtue of (\ref{eqy}) and the fact that $0 \leq \ell < L$, we have
\[
0 \leq \frac{y_1}{L} - z_1 < 1,
\]
which leads to
\[
\left \lfloor \frac{y_1}{L} - z_1 \right \rfloor = 0
\]
and consequently \be z_1 = \left \lfloor \frac{y_1}{L} \right \rfloor. \label{eqa_app_6} \ee Combining
(\ref{eqy}), (\ref{eqa_app_4}), (\ref{eqa_app_5}) and (\ref{eqa_app_6}) yields
\begin{eqnarray*}
y_m & = & \left \lceil   \frac{\varphi_m}{L} - \left (
\frac{y_1}{L} -  z_1 \right )
\; \lambda_m - \frac{1}{2}  \right \rceil - z_1 \; \lambda_m\\
& = & z_m - z_1 \lambda_m \\
& = & z_m - \frac{y_1 - \ell}{L} \lambda_m
\end{eqnarray*}
for $m = 2, \cdots, M$.  This proves (\ref{eqa_app_3}).  It
follows that (\ref{eqyms}) is true and the lemma is thus proven.

\end{pf}

\bigskip

\begin{lemma} \label{right}
Let $y = [y_1, \;\cdots, y_M]  \in \mathcal{ S}$.  If $\widetilde{\ell} = y_1 -  \left \lfloor \frac{y_1}{L}
\right \rfloor L$, then $0 \leq \widetilde{\ell} < L$ and \be \sum_{m =1}^M [(C_m \lambda_m \; \widetilde{\ell}
- C_m \varphi_m) \; {\rm mod}^* C_m L]^2 = || y \; G - \xi ||^2. \label{equ2} \ee
\end{lemma}

\bigskip

\begin{pf}
By the definition of $\widetilde{\ell}$,
\[
\frac{ \widetilde{\ell} } {L} = \frac{y_1}{L} - \left \lfloor
\frac{y_1}{L} \right \rfloor \in [0,1).
\]
Hence, $0 \leq \widetilde{\ell} < L$.  Clearly, there uniquely exist integers $\widetilde{z}_1, \cdots,
\widetilde{z}_M$ such that, for $m = 1, \cdots, M$,
\[
(C_m \lambda_m \; \widetilde{\ell} - C_m \varphi_m) \; {\rm mod}^* C_m L = C_m \lambda_m \; \widetilde{\ell} +
\widetilde{z}_m C_m L - C_m \varphi_m
\]
where $\lambda_1 = 1$.  Therefore,  to prove (\ref{equ2}) it suffices to show \be C_1 \; \widetilde{\ell} +
\widetilde{z}_1 C_1 L - C_1 \varphi_1 = C_1 y_1 - C_1 \varphi_1 \label{eqlem21} \ee and, for $m = 2, \cdots, M$,
\be C_m \lambda_m \; \widetilde{\ell} + \widetilde{z}_m C_m L - C_m \varphi_m =  C_m \lambda_m \; y_1 + C_m L \;
y_m - C_m \varphi_m. \label{eqlem22} \ee Equation (\ref{eqlem21}) can be simplified as \be \widetilde{\ell} +
\widetilde{z}_1 L = y_1, \label{lem2_0} \ee which can be further reduced to \be \widetilde{z}_1 = \left \lfloor
\frac{y_1}{L} \right \rfloor \label{eqlem23} \ee by invoking the definition of $\widetilde{\ell}$.  By the
definition of the symmetric modulus operator ${\rm mod}^*$,
\[
- \frac{C_1 L}{2} \leq C_1  \; \widetilde{\ell} + \widetilde{z}_1
C_1 L - C_1 \varphi_1 < \frac{C_1 L}{2},
\]
which can be rewritten as
\[
-1 < \left( \frac{ \varphi_1 - \widetilde{\ell} }{L} - \frac{1}{2}
\right) - \widetilde{z}_1 \leq 0,
\]
or equivalently, \be \left \lceil \left( \frac{ \varphi_1 -
\widetilde{\ell} }{L} - \frac{1}{2} \right)  - \widetilde{z}_1
\right \rceil = 0. \label{lem2_1} \ee By virtue of (\ref{lem2_1})
and the definition of $\widetilde{\ell}$,
\begin{eqnarray}
\widetilde{z}_1 & = & \left \lceil \frac{ \varphi_1 -
\widetilde{\ell} }{L}
- \frac{1}{2} \right \rceil \nonumber\\
& = & \left \lceil \frac{ \varphi_1 - ( y_1 -  \left \lfloor
\frac{y_1}{L} \right \rfloor L ) }{L} - \frac{1}{2}
\right \rceil \nonumber\\
& = & \left \lceil  \frac{\varphi_1}{L} - \frac{y_1}{L} -
\frac{1}{2} \right \rceil + \left \lfloor \frac{y_1}{L} \right
\rfloor. \label{lem2_2}
\end{eqnarray}
Remember that $y_1$ is restricted by condition
\[
- \frac{L}{2} + \varphi_1 \leq y_1 < \frac{L}{2} + \varphi_1,
\]
or equivalently
\[
-1 < \frac{\varphi_1}{L} - \frac{y_1}{L} - \frac{1}{2}  \leq 0
\]
which implies \be \left \lceil  \frac{\varphi_1}{L} -
\frac{y_1}{L} - \frac{1}{2} \right \rceil = 0. \label{lem2_3} \ee
Combining (\ref{lem2_2}) with (\ref{lem2_3}) yields
(\ref{eqlem23}) and consequently proves (\ref{eqlem21}).

\bigskip

We now turn our attention to the proof of (\ref{eqlem22}). By the
definition of $\widetilde{\ell}$,
 (\ref{eqlem22}) can be rewritten as
\[
\lambda_m  \left(  y_1 - \left \lfloor \frac{y_1}{L} \right
\rfloor L \right) + \widetilde{z}_m L = y_1 \lambda_m + y_m L,
\]
which can be further simplified as \be \widetilde{z}_m  = y_m +
\left \lfloor \frac{y_1}{L} \right \rfloor \lambda_m.
\label{eqlem25} \ee By the definition of the symmetric modulus
operator ${\rm mod}^*$, we have
\[
- \frac{C_m L}{2} \leq C_m \lambda_m \; \widetilde{\ell} +
\widetilde{z}_m C_m L - C_m \varphi_m < \frac{C_m L}{2},
\]
which can be rewritten as
\[
- \frac{C_m L}{2} \leq C_m [\lambda_m (  \widetilde{\ell} + \widetilde{z}_1 L ) + ( \widetilde{z}_m  -
\widetilde{z}_1 \lambda_m) L - \varphi_m] < \frac{C_m L}{2},
\]
or equivalently,
\[
-1 < \frac{ \varphi_m  } {L} - \left(  \frac{\widetilde{\ell} +
\widetilde{z}_1 L } { L} - \widetilde{z}_1 \right)  \lambda_m -
\frac{1}{2}  - \widetilde{z}_m \leq 0.
\]
Hence,
\[
\left \lceil  \frac{ \varphi_m  } {L} - \left(
\frac{\widetilde{\ell} + \widetilde{z}_1 L } { L} -
\widetilde{z}_1 \right ) \lambda_m - \frac{1}{2}  -
\widetilde{z}_m  \right \rceil = 0.
\]
Since $\widetilde{z}_m$ is an integer, it can be determined that \be \widetilde{z}_m  = \left \lceil  \frac{
\varphi_m  } {L} - \left(  \frac{\widetilde{\ell} + \widetilde{z}_1 L } { L} - \widetilde{z}_1 \right )
\lambda_m - \frac{1}{2}  \right \rceil. \label{lem2_4} \ee Note that (\ref{lem2_0}) is true since
(\ref{eqlem23}) has been established.  Using (\ref{lem2_0}), (\ref{eqlem23}) and (\ref{lem2_4}), we obtain \be
\widetilde{z}_m = \left \lceil  \frac{ \varphi_m  } {L} - \left(  \frac{y_1} { L} -   \left \lfloor
\frac{y_1}{L} \right \rfloor \right ) \lambda_m - \frac{1}{2}  \right \rceil. \label{lem2_5} \ee Invoking
(\ref{eqa_app_5}) and (\ref{lem2_5}) leads to (\ref{eqlem25}). This proves (\ref{eqlem22}) and the proof of the
lemma is thus completed.

\end{pf}

\bigskip

We are now in position to prove Theorem~\ref{reduce}.  By
Lemma~\ref{left}, we have
\[
\min_{\ell} \; \sum_{m =1}^M [(C_m \lambda_m \; \ell - C_m \varphi_m) \; {\rm mod}^* C_m L]^2 \;  \geq \;
\min_{y \in \mathcal{ S} } \; || y G - \xi ||^2.
\]
On the other hand, by Lemma~\ref{right}, we have
\[
\min_{\ell} \; \sum_{m =1}^M [(C_m \lambda_m \; \ell - C_m \varphi_m) \; {\rm mod}^* C_m L]^2 \; \leq  \;
\min_{y \in \mathcal{ S} } \; || y G - \xi ||^2.
\]
Therefore,
\[
\min_{\ell} \; \sum_{m =1}^M [(C_m \lambda_m \; \ell - C_m \varphi_m) \; {\rm mod}^* C_m L]^2  \; =  \; \min_{y
\in \mathcal{ S} } \; || y G - \xi ||^2.
\]
Since $\widehat{y}_1$ is the first entry of
\[
\widehat{y} = [\widehat{y}_1, \cdots, \widehat{y}_M] \; = \; \arg \; \min_{y \in \mathcal{ S} } \; || y G - \xi
||^2
\]
and $\widehat{\ell}$ is unique, it follows from Lemma~\ref{right}
that
\begin{eqnarray*}
\widehat{\ell} & = & \arg  \; \min_{\ell} \; \sum_{m =1}^M [(C_m \lambda_m \; \ell - C_m \varphi_m) \; {\rm
mod}^* C_m L]^2\\
 & = & \; \widehat{y}_1 - \left \lfloor \frac{ \widehat{y}_1 } { L } \right \rfloor \; L.
\end{eqnarray*}
The proof of Theorem~\ref{reduce} is thus completed.

\bigskip

\section{PROOF OF THEOREM \ref{th_bbb}} \label{app4}

By the definitions of $G$ and $\xi$,
\[
 || y G - \xi||^2 =
[ C_1 (y_1 - \varphi_1) ]^2 + \sum_{m=2}^M  [ C_m (\lambda_m \;
y_1 + L \; y_m - \varphi_m )]^2.
\]
Note that
\[
[ C_1 (y_1 - \varphi_1) ]^2 + \sum_{m=2}^M  [ C_m (\lambda_m \;
y_1 + L \; y_m - \varphi_m )]^2 < \gamma^2
\]
if and only if
\begin{eqnarray}
&   & [ C_1 (y_1 - \varphi_1) ]^2 < \gamma^2,
\label{equp}\\
&   & [ C_1 (y_1 - \varphi_1) ]^2 + \sum_{p=2}^m [ C_p (\lambda_p \; y_1 + L \; y_p - \varphi_p )]^2 < \gamma^2
\nonumber\\
&    & \;\;\;\; {\rm for} \;\;\;\; m = 2, \cdots,M. \label{eqlow}
\end{eqnarray}
Inequalities (\ref{equp}) and (\ref{eqlow}) can be shown to be equivalent to (\ref{boundary0}) and
(\ref{boundary}) by invoking the definitions of $\mathcal{ S}$ and $\mu_m$.

\bigskip

\section{DATA OF UNITARY SPACE-TIME CODES}  \label{app5}

\bigskip

\underline{Constellation for $M = N = b = 2, \; L = 1024, \; R =
6$}

\bigskip
\bigskip

For $q = 0, 1, 2, 3$,
\[
\Lambda_q  = {\rm diag} \left( \exp \left( \frac{2 \pi i }{L} [1
\;\;  376] \right) \right),
\]
\[
A_q = B_0  = I_{2 \times 2}.
\]
\bigskip

\[
B_1  = \left[\begin{array}{rr}
0.5192 + 0.1730i   & 0.7689 + 0.3305i\\
   0.3249 + 0.7713i  & -0.1692 - 0.5205i
\end{array} \right],
\]
\[
B_2  = \left[\begin{array}{rr}
0.4772 - 0.3219i   & 0.0907 + 0.8127i\\
  -0.1774 + 0.7983i  & 0.4398 + 0.3713i
\end{array} \right],
\]
\[
B_3 = \left[\begin{array}{rr}
-0.4458 + 0.3772i  & 0.7645 + 0.2729i\\
  -0.6303 + 0.5115i & -0.5459 - 0.2075i
\end{array} \right].
\]

\bigskip

See Figure \ref {fig18} for the corresponding performance
simulation results.

\bigskip

\bigskip

\underline{Constellation for $M = N = 2, \; b = 3, \; L = 512, \;
R = 6$}
\bigskip
\bigskip

For $q = 0, 1, \cdots , 7$,
\[
\Lambda_q  = {\rm diag} \left( \exp \left( \frac{2 \pi i }{L} [1
\;\;  188] \right) \right), \]
\[
A_q = B_0  = I_{2 \times 2}.
\]
\bigskip
\[
B_1  = \left[\begin{array}{rr}
0.3408 + 0.6630i & -0.1400 - 0.6517i\\
  -0.2401 + 0.6218i  & -0.4402 + 0.6016i\end{array} \right],
\]
\[
B_2  = \left[\begin{array}{rr}
0.4230 + 0.2881i &  -0.7279 + 0.4563i\\
   0.8585 + 0.0319i  & 0.2226 - 0.4609i \end{array} \right],
\]
\[
B_3 = \left[\begin{array}{rr}
0.3663 + 0.1357i  & 0.8257 + 0.4069i\\
  -0.5379 - 0.7470i  & 0.1944 + 0.3388i \end{array} \right],
\]
\[
B_4 = \left[\begin{array}{rr}
   0.7428 + 0.1845i  & 0.4221 - 0.4858i\\
  -0.6130 - 0.1962i  & 0.5391 - 0.5433i
\end{array} \right],
\]
\[
B_5 = \left[\begin{array}{rr}
   0.2656 - 0.1927i & -0.9238 + 0.1975i\\
   0.3009 + 0.8954i  & -0.0304 + 0.3267i
\end{array} \right],
\]
\[
B_6 = \left[\begin{array}{rr}
   0.0816 + 0.8219i & -0.4396 - 0.3530i\\
   0.3081 + 0.4722i  & 0.8099 + 0.1620i
\end{array} \right],
\]
\[
B_7 = \left[\begin{array}{rr}
  -0.0442 - 0.7407i  &  0.5677 - 0.3564i\\
  -0.5320 - 0.4079i &  -0.1131 + 0.7334i
  \end{array} \right].
\]

\bigskip

See Figure \ref {fig18} for the corresponding performance
simulation results.

\bigskip

\underline{Constellation for $M = N = 2, \; b = 4, \; L = 256, \;
R = 6$}

\bigskip

\bigskip

For $q = 0, 1, \cdots , 15$,
\[
\Lambda_q  = {\rm diag} \left( \exp \left( \frac{2 \pi i }{L} [1
\;\;  75.7044] \right) \right),
\]
\[
A_q = B_0 = I_{2 \times 2}.
\]

\bigskip

\[
B_1  = \left[\begin{array}{rr}
0.3912 - 0.8587i  & 0.1204 - 0.3083i\\
0.2004 + 0.2635i  & -0.6117 - 0.7185i
\end{array} \right],
\]
\[
B_2  = \left[\begin{array}{rr}
-0.1412 + 0.1279i  & 0.1820 + 0.9647i\\
0.7979 - 0.5718i  & 0.0138 + 0.1900i
\end{array} \right],
\]
\[
B_3 = \left[\begin{array}{rr}
0.4099 + 0.6855i &  0.5015 - 0.3325i\\
  -0.5988 - 0.0590i  &  0.0412 - 0.7976i
\end{array} \right],
\]
\[
B_4 = \left[\begin{array}{rr}
0.2787 + 0.3877i  &  -0.6636 - 0.5759i\\
0.8235 + 0.3064i  &  0.4739 + 0.0588i
\end{array} \right],
\]
\[
B_5 = \left[\begin{array}{rr}
-0.7060 + 0.4436i & -0.3287 - 0.4435i\\
   0.5057 - 0.2215i  &  -0.3922 - 0.7358i
\end{array} \right],
\]
\[
B_6 = \left[\begin{array}{rr}
0.5580 + 0.0014i  & -0.0343 - 0.8292i\\
0.7699 - 0.3096i  &  0.2307 + 0.5080i
\end{array} \right],
\]
\[
B_7 = \left[\begin{array}{rr}
-0.2027 + 0.7271i  &  0.2565 + 0.6037i\\
  -0.6504 - 0.0851i  & 0.6461 - 0.3904i
\end{array} \right],
\]
\[
B_8 = \left[\begin{array}{rr}
-0.2526 + 0.2877i  & -0.2635 - 0.8854i\\
   0.3749 - 0.8443i  & -0.2137 - 0.3177i
\end{array} \right],
\]
\[
B_9 = \left[\begin{array}{rr}
-0.8735 + 0.1501i  &  0.2828 - 0.3666i\\
   0.4626 + 0.0203i  &  0.6780 - 0.5708i
\end{array} \right],
\]
\[
B_{10} = \left[\begin{array}{rr}
-0.8175 - 0.4231i  & -0.0195 + 0.3903i\\
   0.2927 - 0.2589i  &  0.8409 + 0.3744i
\end{array} \right],
\]
\[
B_{11} = \left[\begin{array}{rr}
-0.5913 - 0.4068i  & -0.4858 - 0.4989i\\
  -0.6674 - 0.1988i  &  0.6347 + 0.3350i\\
\end{array} \right],
\]
\[
B_{12} = \left[\begin{array}{rr}
0.3609 + 0.1064i  &  0.7232 + 0.5792i\\
   0.9174 + 0.1295i  & -0.3248 - 0.1899i
\end{array} \right],
\]
\[
B_{13} = \left[\begin{array}{rr}
-0.1464 - 0.8164i  &  -0.0346 + 0.5575i\\
  -0.4076 + 0.3819i  & -0.7225 + 0.4075i
\end{array} \right],
\]
\[
B_{14} = \left[\begin{array}{rr}
-0.0575 + 0.6282i  & -0.6849 + 0.3647i\\
   0.3285 + 0.7030i  &  0.3313 - 0.5368i
\end{array} \right],
\]
\[
B_{15} = \left[\begin{array}{rr}
0.4912 + 0.3594i  & -0.0364 - 0.7926i\\
   0.3456 + 0.7142i  & -0.2606 + 0.5500i
\end{array} \right].
\]

\bigskip
\bigskip

See Figure \ref {fig18} for the corresponding performance simulation results.


\underline{Constellation for $M =3, \;  N = 1, \; b = 4, \;L =
256, \; R = 4$ (see Figure \ref {fig28})}

\[
\Lambda_q  = {\rm diag} \left( \exp \left( \frac{2 \pi i }{L} [1
\;\;  33.7365 \;\;58.5425] \right) \right), \quad A_q = B_0  =
I_{3 \times 3}, \;\;\;\;\;\;  q = 0, 1, \cdots, 15.
\]
\[
B_{1} = \left[\begin{array}{rrr}
   0.7602 + 0.1419i  &  -0.3318 - 0.3072i   &  0.2330 + 0.3785i\\
  -0.0629 - 0.2186i  &  -0.1319 - 0.8171i  &  -0.1144 - 0.5002i\\
   0.2379 + 0.5419i  &  -0.1797 + 0.2798i   &  0.1680 - 0.7148i
\end{array} \right],
\]
\[
B_{2} = \left[\begin{array}{rrr}
-0.1626 + 0.2936i  &  -0.2426 - 0.5136i   &  0.2881 + 0.6941i\\
  -0.2142 + 0.8692i   &  0.3728 + 0.1549i   &  0.0315 - 0.1861i\\
   0.0007 + 0.2931i  &  -0.6903 + 0.1948i  &  -0.6309 + 0.0408i
\end{array} \right],
\]
\[
B_{3} = \left[\begin{array}{rrr}
  -0.3998 + 0.6106i  &  -0.6134 + 0.2661i   &  0.0346 - 0.1385i\\
   0.0378 - 0.1693i  &  -0.2662 - 0.0232i  &  -0.9470 + 0.0426i\\
   0.5614 - 0.3495i  &  -0.6931 + 0.0357i   &  0.2809 + 0.0468i
\end{array} \right],
\]
\[
B_{4} = \left[\begin{array}{rrr}
  -0.0076 + 0.1851i   &  0.2440 - 0.9364i  &  -0.1497 - 0.0827i\\
   0.2850 - 0.6178i  &  -0.0305 - 0.0343i  &  -0.2300 - 0.6943i\\
   0.2799 - 0.6515i   &  0.1769 - 0.1737i   &  0.2094 + 0.6260i
\end{array} \right],
\]
\[
B_{5} = \left[\begin{array}{rrr}
  -0.0148 + 0.0411i   &  0.5416 - 0.0737i   &  0.1535 - 0.8221i\\
   0.2580 - 0.1099i  &  -0.5039 + 0.6258i  &  -0.2723 - 0.4490i\\
   0.9480 - 0.1438i   &  0.2031 - 0.1203i  &  0.0806 + 0.1354i
\end{array} \right],
\]
\[
B_{6} = \left[\begin{array}{rrr}
   0.6097 + 0.1555i   &  0.0383 + 0.3146i  &  -0.5723 + 0.4197i\\
   0.4287 - 0.0217i  &  -0.3612 + 0.5689i   &  0.5271 - 0.2896i\\
   0.6463 + 0.0458i  &   0.1922 - 0.6392i   &  0.1045 - 0.3518i
\end{array} \right],
\]
\[
B_{7} = \left[\begin{array}{rrr}
  -0.0706 - 0.1161i   &  0.1815 + 0.0717i   &  0.8110 + 0.5345i\\
  -0.4231 + 0.4203i   &  0.7746 + 0.1128i  &  -0.1761 + 0.0243i\\
   0.3546 - 0.7072i   &  0.5745 - 0.1383i  &  -0.1161 - 0.1071i
\end{array} \right],
\]
\[
B_{8} = \left[\begin{array}{rrr}
   0.2373 + 0.6258i   &  0.6446 - 0.3587i  &  -0.0844 - 0.0277i\\
  -0.0899 - 0.5506i   &  0.5100 - 0.0798i  &  -0.2909 + 0.5811i\\
  -0.2728 + 0.4080i  &  -0.4228 - 0.1028i  &  -0.6564 + 0.3728i
\end{array} \right],
\]
\[
B_{9} = \left[\begin{array}{rrr}
   0.2476 + 0.3168i  &  -0.3066 - 0.6917i  &  -0.3383 - 0.3891i\\
   0.4145 - 0.3141i   &  0.6045 - 0.2448i  &  -0.4322 + 0.3426i\\
   0.5080 - 0.5566i  &  -0.0430 + 0.0190i   &  0.4250 - 0.4993i
\end{array} \right],
\]
\[
B_{10} = \left[\begin{array}{rrr}
  -0.4832 - 0.0060i  &  -0.4217 + 0.6427i   &  0.4049 + 0.1075i\\
   0.2751 - 0.0456i   &  0.2416 - 0.2205i   &  0.8445 + 0.3195i\\
  -0.1269 - 0.8201i   &  0.4829 + 0.2623i  &  -0.0876 + 0.0397i
\end{array} \right],
\]
\[
B_{11} = \left[\begin{array}{rrr}
   0.3610 + 0.4449i   &  0.2997 + 0.4246i   &  0.5036 + 0.3846i\\
  -0.2332 + 0.7337i   &  0.1987 - 0.0629i  &  -0.0908 - 0.5963i\\
  -0.2399 + 0.1465i   &  0.4152 - 0.7170i   &  0.0309 + 0.4833i
\end{array} \right],
\]
\[
B_{12} = \left[\begin{array}{rrr}
  -0.0346 - 0.0609i  &  -0.6038 + 0.5991i  &  -0.4153 - 0.3148i\\
   0.0784 + 0.8724i  &  -0.1136 + 0.2598i   &  0.3898 + 0.0206i\\
  -0.3106 - 0.3625i   &  0.1769 + 0.4060i   &  0.6732 - 0.3505i
\end{array} \right],
\]
\[
B_{13} = \left[\begin{array}{rrr}
  -0.0800 - 0.3611i  &  -0.4150 + 0.0674i  &  -0.0114 - 0.8284i\\
  -0.0151 + 0.0077i  &  -0.7741 + 0.4418i   &  0.1714 + 0.4195i\\
  -0.3292 - 0.8687i   &  0.1661 - 0.0354i   &  0.0598 + 0.3235i
\end{array} \right],
\]
\[
B_{14} = \left[\begin{array}{rrr}
  -0.0408 + 0.0395i   &  0.4025 + 0.4897i  &  -0.1679 - 0.7529i\\
   0.8274 - 0.2735i   &  0.3768 + 0.1065i   &  0.2514 + 0.1554i\\
   0.4822 + 0.0702i  &  -0.5340 - 0.3997i  &  -0.2202 - 0.5188i
\end{array} \right],
\]
\[
B_{15} = \left[\begin{array}{rrr}
   0.4458 + 0.2691i  &  -0.7835 + 0.2801i  &  -0.1008 - 0.1623i\\
  -0.5050 - 0.1408i  &  -0.4301 + 0.1993i   &  0.1919 + 0.6809i\\
   0.1380 - 0.6595i  &  -0.2277 - 0.1762i   &  0.6115 - 0.2987i
\end{array} \right].
\]

\bigskip

\underline{Constellation for $M =4, \;  N = 2,\; b = 2, \; L = 64,
\; R = 2$ (see Figure \ref {fig38})}

\bigskip

\[
\Lambda_q  = {\rm diag} \left( \exp \left( \frac{2 \pi i }{L} [1
\;\; 5 \;\; 17 \;\; 28 ] \right) \right),
  \;\;\; \;\;\; A_q = B_0  = I_{4 \times 4}, \;\;\;\;  q = 0, 1, 2, 3.
\]
\[
B_1  = \left[\begin{array}{rrrr}
   0.1920 - 0.0840i  & -0.2404 - 0.0482i   & 0.4479 - 0.5434i  & -0.5535 -
0.3061i\\
   0.2506 - 0.2836i  & -0.0749 + 0.0316i  & -0.3375 + 0.5975i  & -0.5664 -
0.2417i\\
   0.5003 + 0.4453i  & -0.3540 - 0.4598i  & -0.0062 + 0.1056i  & -0.0564 +
0.4475i\\
  -0.5925 + 0.1151i  & -0.7710 - 0.0446i  & -0.1140 + 0.0945i  & 0.0034 -
0.1313i
\end{array} \right],
\]
\[
B_2  = \left[\begin{array}{rrrr}
   0.0820 + 0.1057i  & -0.3400 - 0.6390i  & -0.1703 - 0.4908i   & 0.3557 -
0.2486i\\
  -0.1493 - 0.5629i  &  0.0330 + 0.1103i  & -0.0758 + 0.3064i   & 0.7293 -
0.1268i\\
   0.4421 + 0.5263i  &  0.2416 + 0.4456i  & -0.0185 - 0.1039i   & 0.4068 -
0.3064i\\
  -0.3881 - 0.1410i   & 0.3614 + 0.2743i  & -0.5385 - 0.5739i  & -0.0546 +
0.0366i\\
\end{array} \right],
\]
\[
B_3 = \left[\begin{array}{rrrr}
  -0.0069 + 0.0651i   & 0.5578 + 0.1900i  & -0.2629 - 0.3439i   & 0.2290 -
0.6392i\\
   0.3545 - 0.6450i  &  -0.1827 - 0.2323i  &  0.4409 - 0.1617i  &  0.0128 -
0.3878i\\
  -0.1216 + 0.4867i  & -0.4988 + 0.1761i   & 0.2493 + 0.1850i  & -0.1557 -
0.5899i\\
   0.1847 - 0.4103i  & -0.1855 + 0.5012i  & -0.4816 + 0.5137i  &  0.0359 -
0.1220i

\end{array} \right].
\]

\bigskip

\underline{Constellation for $M =4, \;  N = 2, \; b = 4, \; L =
256, \; R = 3$ (see Figure \ref {fig48})}

\bigskip

\[
\Lambda_q  = {\rm diag} \left( \exp \left( \frac{2 \pi i }{L} [1
\;\; 7.9761 \;\; 68.6816 \;\; 106.6000] \right) \right),
 \;\;\; A_q = B_0  = I_{4 \times 4}, \;\;\;  q = 0, 1, \cdots, 15.
\]
\[
B_1  = \left[\begin{array}{rrrr}
  -0.4860 - 0.2228i  &  -0.6620 - 0.2202i  &  -0.1202 - 0.3650i   &  0.0005
+ 0.2823i\\
   0.2336 - 0.2148i   &  0.2808 + 0.1490i  &  -0.1188 - 0.6816i   &  0.5604
+ 0.0744i\\
   0.0589 + 0.3944i  &  -0.0909 + 0.0457i   &  0.0637 - 0.5901i  &  -0.4779
- 0.5000i\\
   0.6445 - 0.1977i  &  -0.6320 + 0.0494i  &  -0.0009 + 0.1460i   &  0.1766
- 0.3019i
\end{array} \right],
\]
\[
B_{2}  = \left[\begin{array}{rrrr}
  -0.4407 + 0.1717i  &  -0.6272 + 0.0508i   &  0.4816 + 0.1003i  &  -0.2527
+ 0.2729i\\
  -0.8127 - 0.0589i   &  0.4192 + 0.0524i  &  -0.0967 - 0.3587i  &  -0.0923
- 0.1050i\\
   0.0371 - 0.3235i  &  -0.3105 - 0.1548i  &  -0.5337 - 0.4044i  &  -0.1209
+ 0.5573i\\
   0.0490 - 0.0620i  &  0.5423 + 0.1058i   &  0.1616 + 0.3814i  &  -0.2759 +
0.6639i
\end{array} \right],
\]
\[
B_{3}  = \left[\begin{array}{rrrr}
  -0.2059 - 0.1255i   &  0.0653 + 0.2605i  &  -0.7272 + 0.4766i  &  -0.0893
- 0.3254i\\
  -0.5978 - 0.0094i   &  0.0292 - 0.0758i   &  0.3068 - 0.0283i   &  0.5017
- 0.5378i\\
  -0.0554 - 0.1514i   &  0.3235 - 0.8910i  &  -0.2501 - 0.0408i  &  -0.1019
- 0.0281i\\
  -0.1131 - 0.7386i  &  -0.1193 + 0.0915i   &  0.2566 - 0.1384i  &  -0.5327
- 0.2241i
\end{array} \right],
\]
\[
B_{4} = \left[\begin{array}{rrrr}
   0.2059 - 0.4636i   &  0.6366 + 0.2002i   &  0.1927 - 0.2336i  &  -0.3935
- 0.2254i\\
   0.2157 - 0.2034i   &  0.0779 + 0.1506i  &  -0.0209 + 0.7166i  &  -0.2077
+ 0.5712i\\
  -0.1516 - 0.5863i  &  -0.3966 - 0.2471i   &  0.0283 + 0.3719i  &  -0.0641
- 0.5212i\\
   0.5363 + 0.0211i  &  -0.4133 - 0.3698i   &  0.4294 - 0.2661i  &  -0.3389
+ 0.1852i
\end{array} \right],
\]
\[
B_{5}  = \left[\begin{array}{rrrr}
   0.1610 - 0.0104i  &  -0.1781 - 0.0118i  &  -0.3241 + 0.4502i   &  0.5883
- 0.5370i\\
  -0.1163 - 0.7159i  &  -0.4619 + 0.0427i  &  -0.3587 + 0.1576i  &  -0.2228
+ 0.2359i\\
   0.4255 + 0.4822i  &  -0.1864 + 0.4173i  &  -0.2767 + 0.3474i  &  -0.3281
+ 0.2696i\\
  -0.0127 + 0.1852i  &  -0.3440 - 0.6525i   &  0.3753 + 0.4480i   &  0.0114
+ 0.2825i
\end{array} \right],
\]
\[
B_{6} = \left[\begin{array}{rrrr}
   0.5947 + 0.0898i   &  0.0777 - 0.0544i   &  0.3469 - 0.1757i  &  -0.6284
+ 0.2883i\\
   0.3998 + 0.2454i   &  0.1581 - 0.4791i   &  0.2068 + 0.3971i   &  0.5486
+ 0.1550i\\
   0.6131 - 0.1431i   &  0.2198 + 0.2581i  &  -0.6078 + 0.0618i   &  0.0583
- 0.3347i\\
   0.1429 + 0.0378i  &  -0.6927 + 0.3764i  &  -0.0616 + 0.5208i  &  -0.0268
+ 0.2844i
\end{array} \right],
\]
\[
B_{7}  = \left[\begin{array}{rrrr}
   0.5078 + 0.1352i   &  0.0293 - 0.6221i  &  -0.3714 - 0.0843i   &  0.1771
- 0.3996i\\
  -0.0888 + 0.0556i  &  -0.0162 + 0.1379i  &  -0.8660 - 0.0570i  &  -0.1509
+ 0.4401i\\
   0.4760 - 0.1213i  &  -0.0101 - 0.3142i   &  0.2954 - 0.0688i  &  -0.1032
+ 0.7465i\\
  -0.6725 - 0.1394i  &   0.1043 - 0.6951i   &  0.0136 + 0.0976i  &  -0.1192
+ 0.1021i
\end{array} \right],
\]
\[
B_{8} = \left[\begin{array}{rrrr}
   0.2398 - 0.2323i   &  0.2102 + 0.2519i  &  -0.0454 + 0.7652i   &  0.4128
+ 0.1513i\\
  -0.4485 + 0.0876i   &  0.4478 + 0.3515i  &  -0.2440 - 0.3668i   &  0.5225
- 0.0054i\\
   0.2040 - 0.3878i   &  0.6345 + 0.2146i  &  -0.1047 - 0.1151i  &  -0.5780
- 0.0302i\\
   0.5155 - 0.4710i  &  -0.0559 - 0.3413i  &  -0.1917 - 0.3967i   &  0.4298
- 0.1182i
\end{array} \right],
\]
\[
B_{9}  = \left[\begin{array}{rrrr}
  -0.0808 + 0.1747i   &  0.6974 + 0.0254i  &  -0.3297 + 0.1999i  &  -0.4411
+ 0.3643i\\
  -0.2270 - 0.1530i   &  0.3884 + 0.1536i   &  0.2603 + 0.6090i   &  0.2922
- 0.4760i\\
  -0.5193 + 0.7098i  &  -0.3679 + 0.1068i  &  -0.0637 + 0.2636i   &  0.0287
+ 0.0737i\\
   0.0413 - 0.3359i  &  -0.3492 + 0.2643i   &  0.3891 + 0.4333i  &  -0.4673
+ 0.3690i
\end{array} \right],
\]
\[
B_{10} = \left[\begin{array}{rrrr}
   0.6160 + 0.1969i  &  -0.4480 - 0.4219i   &  0.1428 - 0.3612i  &  -0.2164
- 0.0733i\\
   0.2687 + 0.0204i   &  0.6471 + 0.3236i  &  -0.0287 - 0.6065i  &  -0.1607
- 0.0976i\\
   0.0571 - 0.4298i   &  0.0075 - 0.1040i   &  0.0461 + 0.0462i   &  0.2063
- 0.8685i\\
   0.5006 - 0.2658i   &  0.2913 + 0.0464i   &  0.3550 + 0.5917i  &  -0.3108
+ 0.1379i
\end{array} \right],
\]
\[
B_{11}  = \left[\begin{array}{rrrr}
   0.7748 - 0.0926i  &  -0.1322 - 0.1752i   &  0.0966 + 0.3047i   &  0.2966
+ 0.3909i\\
  -0.2424 - 0.3439i   &  0.3463 - 0.1798i   &  0.0991 + 0.7302i  &  -0.3426
+ 0.1017i\\
   0.2561 - 0.1187i  &  -0.3777 - 0.5443i   &  0.0569 - 0.1325i  &  -0.5445
- 0.4052i\\
   0.2411 - 0.2762i   &  0.6003 + 0.0157i   &  0.2016 - 0.5416i  &  -0.3096
+ 0.2740i
\end{array} \right],
\]
\[
B_{12}  = \left[\begin{array}{rrrr}
   0.1280 + 0.2465i  &  -0.6345 - 0.1417i  &  -0.0766 - 0.4249i   &  0.0486
- 0.5581i\\
   0.0816 - 0.6158i   &  0.2517 - 0.0380i   &  0.0347 + 0.2233i   &  0.1347
- 0.6929i\\
   0.2862 - 0.1410i  &  -0.3924 + 0.2196i   &  0.7920 + 0.2295i  &  -0.0725
+ 0.1040i\\
  -0.2304 - 0.6181i  &  -0.5369 - 0.1485i  &  -0.2868 + 0.0181i   &  0.1966
+ 0.3650i
\end{array} \right],
\]
\[
B_{13}  = \left[\begin{array}{rrrr}
   0.0256 - 0.1183i   &  0.1067 + 0.2697i   &  0.1343 - 0.1014i   &  0.9325
- 0.0583i\\
  -0.5312 - 0.2629i  &  -0.6552 - 0.4068i  &  -0.0562 - 0.1474i   &  0.1628
- 0.0512i\\
   0.5708 - 0.4629i  &  -0.4947 + 0.2483i   &  0.1888 + 0.1781i  &  -0.1143
- 0.2703i\\
   0.0258 + 0.3055i  &  -0.0856 - 0.0859i   &  0.8342 - 0.4284i  &  -0.0972
- 0.0495i
\end{array} \right],
\]
\[
B_{14}  = \left[\begin{array}{rrrr}
   0.3080 - 0.2933i   &  0.4685 - 0.1297i   &  0.2442 - 0.6850i   &  0.1051
- 0.2072i\\
   0.0902 - 0.4068i   &  0.0109 + 0.4103i   &  0.0850 + 0.1134i  &  -0.7756
- 0.1903i\\
  -0.1509 + 0.2560i   &  0.7121 - 0.2210i  &  -0.0907 + 0.4222i  &  -0.1097
- 0.3966i\\
   0.7444 + 0.0551i   &  0.1963 + 0.0281i  &  -0.4936 + 0.1449i  &  -0.0483
+ 0.3696i
\end{array} \right],
\]
\[
B_{15}  = \left[\begin{array}{rrrr}
   0.2287 + 0.2115i  & 0.4099 - 0.4573i & -0.6287 + 0.2885i &  -0.2128 -
0.0457i\\
  -0.3966 - 0.7004i  & -0.0114 - 0.2050i & -0.4089 - 0.1241i  &  0.2918 +
0.2056i\\
  -0.1676 + 0.2912i   &  0.4021 - 0.4722i  &   0.4131 - 0.1470i  &   0.3944
+ 0.3933i\\
  -0.3760 - 0.0294i  &  -0.1257 - 0.4246i   &  0.1711 - 0.3427i  &  -0.7080
- 0.1175i
\end{array} \right].
\]

\bigskip

\begin{table}
\renewcommand{\arraystretch}{1.3}
\caption{Continuous Diagonal Code $\Lambda = {\rm diag}( \exp(
\frac{2 \pi i}{L} u )  )$ } \label{table_example} \centering
\begin{tabular}{|c||c||c||c|}
\hline
$M$ & $R$ & $L$ & $u$\\
\hline
$2$ & $1$ & $4$ & $[1 \;\; 1.6741]$ \\
\hline
$3$ & $1$ & $8$ & $[1 \;\; 1.9537 \;\; 2.9759]$ \\
\hline
$4$ & $1$ & $16$ & $[1 \;\; 2.9976 \;\; 5.0063 \;\; 6.9979]$ \\
\hline $5$ & $1$ & $32$ & $[1 \;\; 2.8963 \;\; 7.9168 \;\; 12.3396
\;\; 14.1375]$
\\
\hline $6$ & $1$ & $64$ & $[1 \;\; 3.9663 \;\; 5.8291 \;\; 17.8483
\;\; 24.6302
\;\; 26.5638]$ \\
\hline $7$ & $1$ & $128$ & $[1 \;\; 3.9607 \;\; 21.9899 \;\;
31.5332 \;\; 47.3852
\;\; 54.2734 \;\; 60.2040]$ \\
\hline
$2$ & $2$ & $16$ & $[1 \;\; 5.9911]$ \\
\hline
$3$ & $2$ & $64$ & $[1 \;\; 6.8881 \;\; 26.5877]$ \\
\hline
$4$ & $2$ & $256$ & $[1 \;\; 7.9761 \;\; 68.6816 \;\; 106.6000]$ \\
\hline $5$ & $2$ & $1024$ & $[1 \;\; 61.0483 \;\; 100.6309 \;\;
129.7491 \;\;
356.4678]$ \\
\hline $6$ & $2$ & $4096$ & $[1 \;\; 11.8659 \;\; 404.3640 \;\;
592.2112 \;\;
1328.7582 \;\; 1489.9040]$ \\
\hline $7$ & $2$ & $16384$ & $[1 \;\; 300.8485 \;\; 4019.3073 \;\;
5142.8482 \;\;
6816.8842 \;\; 8098.6177 \;\; 8109.4273]$ \\
\hline
\end{tabular}
\end{table}


\section{STRUCTURE OF ORTHOGONAL DESIGNS} \label{app6}

In our simulation of orthogonal designs, the frame length is
chosen as $T \geq M$ and the transmitted signals are determined as
\[
S_0 = \sqrt{\frac{T}{M}} \left [ \begin{array}{l} I_{M \times M}\\
 0_{(T-M) \times M} \end{array} \right ], \quad
S_\tau = V_\tau S_{\tau -1}, \quad \tau = 1, 2, \cdots
\]
where $S_\tau$ is a $T \times M$ matrix, $V_\tau =
\mathcal{G}(z_1, \cdots, z_K)$ is defined by a $T \times T$
orthogonal design $\mathcal{G}$ such that $z_1, \cdots, z_K$ are
mapped from PSK constellations $\mathcal{A}_1, \cdots,
\mathcal{A}_K$. The choice of $T$ depends on the number transmit
antennas. For $M = 2$ we choose $T = 2$ and use the $2 \times 2$
orthogonal design in \cite{Alamouti}.
 For $M =3$ and $4$, we choose $T = 4$ and use the $4 \times 4$
 orthogonal design in \cite{Tirkkonen}. For $M = 5, \; 6, \; 7, \;
8$, we chose $T = 8$ and use the $8 \times 8$ orthogonal design in
\cite{Tirkkonen}. It should be noted that such concatenation
between the complex square orthogonal designs and the differential
unitary space-time modulation scheme has been proposed in
\cite{Ganesan} and \cite{Liang1}. For the spectral efficiency to
be an integer $R$, we use the following PSK constellations
\[
\mathcal{A}_k = \left \{ \frac{1}{\sqrt{K}} \exp \left (j  \frac{2 \pi r}{\lceil \frac{TR}{K} \rceil} \right )
\mid r = 0, 1, \cdots, 2^{ \lceil \frac{TR}{K} \rceil } -1 \right \} \] for $1 \leq k \leq ( TR \; \mathrm{mod}
(K) )$;
\[
\mathcal{A}_k = \left \{ \frac{1}{\sqrt{K}} \exp \left (j  \frac{2 \pi r}{\lfloor \frac{TR}{K} \rfloor} \right )
\mid r = 0, 1, \cdots, 2^{ \lfloor \frac{TR}{K} \rfloor } - 1\right \} \] for $( TR \; \mathrm{mod} (K) ) < k
\leq K$.  For the $\tau$-th time frame, bits of length $TR$ are mapped into $z_k \in \mathcal{A}_k, \; k = 1,
\cdots, K$ by Gray codes.  The decoding problem is to solve the minimization problem \be \label{decode} \arg
\min_{ z_k \in \mathcal{A}_k, \; k = 1, \cdots, K  } \| X_\tau - \mathcal{G}(z_1, \cdots, z_K) X_{\tau -1}
\|_\mathrm{F}^2. \ee
 As demonstrated in \cite{Tarokh4},  by exploiting the special structure of the orthogonal design,
the data symbols $z_k \in \mathcal{A}_k, \; k = 1, \cdots, K$ can
be decoupled and decoded individually from (\ref{decode}).

\section*{Acknowledgment}

The authors wish to thank the anonymous reviewers for their
valuable comments and suggestions.

\end{document}